\RequirePackage{fix-cm}
\documentclass[smallcondensed]{svjour3}       
\smartqed  
\usepackage{graphicx}
\usepackage{natbib}
\usepackage{epsfig}
\usepackage{colortbl}
\usepackage{appendix}
\usepackage{ulem}
\usepackage{lineno}

\usepackage{color}
\usepackage{amsmath,amssymb}

\definecolor{Gray}{gray}{0.85}
\definecolor{ForrestGreen}{rgb}{0.133,0.545,0.133}

\definecolor{colga}{rgb}{0.9, 0.6, 0.2}

\definecolor{sy}{rgb}{0.858, 0.188, 0.478}

\usepackage[dvipsnames]{xcolor}

\definecolor{gc}{rgb}{0.2, 0.6, 0.8}

\journalname{Space Science Reviews}
\begin{document}

\title{Decoding the Pre-Eruptive Magnetic Field Configurations of Coronal Mass Ejections}

\titlerunning{Decoding the Pre-Eruptive Magnetic Field Configurations of CMEs}

\author{S. Patsourakos$^1$ \and A. Vourlidas$^2$ \and T. T\"or\"ok$^3$ \and B. Kliem$^4$ \and  
S. K. Antiochos$^5$ \and V. Archontis$^6$ \and G. Aulanier$^7$ \and X. Cheng$^8$ \and G. Chintzoglou$^9$ \and M.K. Georgoulis$^{10}$  \and L.M. Green$^{11}$ \and J. E. Leake$^5$ \and R. Moore$^{12}$ \and A. Nindos$^1$ \and P. Syntelis$^6$ \and S. L. Yardley$^6$ \and V. Yurchyshyn$^{13}$ \and J. Zhang$^{14}$ }

\authorrunning{S. Patsourakos et al.}

\institute{
$^1$Department of Physics, University of Ioaninna, Ioaninna, Greece \\
\and $^2$Johns Hopkins University Applied Physics Laboratory, Laurel, MD, USA \\
\and $^3$Predictive Science Inc., 9990 Mesa Rim Road, Suite 170, San Diego, CA 92121, USA  \\
\and
$^{4}$ Institute of Physics and Astronomy, University of Potsdam, 14476 Potsdam, Germany\\ 
\and $^5$Goddard Space Flight Center, Greenbelt, MD,  USA\\
\and $^6$ School of Mathematics and Statistics, University of St. Andrews, St. Andrews, UK \\
\and $^7$ LESIA, Observatoire de Paris, Universit\'e PSL, CNRS, Sorbonne Universit\'e, Universit\'e de Paris, 5 place Jules Janssen, 92195 Meudon, France \\ \and $^8$
School of Astronomy and Space Science, Nanjing University, Nanjing 210093, People's Republic of China
\\  \and $^9$ Lockheed Martin Solar and Astrophysics Lab, Palo Alto, California, USA \\ \and $^{10}$ 
Research Center Astronomy and Applied Mathematics, Academy of Athens, Athens, Greece \\  \and $^{11}$ University College London, Mullard Space Science Laboratory, Holmbury St. Mary, Dorking, Surrey RH5 6NT, UK \\ 
\and $^{12}$ NASA/Marshall Space Flight Center, Huntsville, Alabama, USA \\
\and
$^{13}$
Big Bear Solar Observatory, New Jersey Institute of Technology, Big Bear City, California, USA \\
\and $^{14}$ Department of Physics and Astronomy, George Mason University, Fairfax, VA, USA
}

\date{Received: date / Accepted: date}

\maketitle
\begin{abstract}
A clear understanding of the nature of the pre-eruptive magnetic field configurations of Coronal Mass Ejections (CMEs) is required for understanding and eventually predicting solar eruptions. Only two, but seemingly disparate, magnetic configurations are considered viable; namely, sheared magnetic arcades (SMA) and magnetic flux ropes (MFR). They can form via three physical mechanisms (flux emergence, flux cancellation, helicity condensation). Whether the CME culprit is an SMA or an MFR, however, has been strongly debated for thirty years. We formed an International Space Science Institute 
(ISSI) team to address and resolve this issue and report  the outcome here.
We review the status of the field across modeling and observations, identify the open and closed issues, compile lists of SMA and MFR observables to be tested against observations and outline research activities to close the gaps in our current understanding. We propose that the combination of multi-viewpoint multi-thermal coronal observations and multi-height vector magnetic field measurements is the optimal approach for resolving the issue conclusively. We demonstrate the approach using MHD simulations and synthetic coronal images. 

Our key conclusion is that the differentiation of pre-eruptive configurations in terms of SMAs and MFRs seems artificial. Both observations and modeling can be made consistent if the pre-eruptive configuration exists in a hybrid state that is continuously evolving from an SMA to an MFR. Thus, the 'dominant' nature of a given configuration will largely depend on its evolutionary stage (SMA-like early-on, MFR-like near the eruption).

\end{abstract}


\section{Introduction} \label{s:intro}

Coronal Mass Ejections (CMEs) are large-scale expulsions of magnetized coronal plasma into the heliosphere. They represent a key energy release process in the solar corona, and a major driver of space weather. Despite the observations and cataloguing of the properties of thousands of events in the last 40 years, the formation and  nature of their pre-eruptive magnetic configurations are still eluding us. Which magnetic configurations are most prone to destabilize and form a CME and how they arise remain hotly debated questions. 

The pre-eruptive magnetic configuration of CMEs has been modeled either as a sheared magnetic arcade (SMA) or as a magnetic flux rope (MFR). In some theoretical/numerical models, the pre-eruptive configuration is an MFR \citep[e.g.,][]{chen1989,amari2000,torok05} and in some the pre-eruptive configuration is a SMA \citep{Moore&Roumeliotis1992, antiochos1999}.

Both of these structures are capable of containing dipped, sheared field lines above a polarity inversion line (PIL), and so are candidates for the magnetic structure of a filament channel (defined in  Section 1.1). Strong PILs are characterized 
by the concentration of most of  the shear (i.e., non-potentiality) of the magnetic field, and hence free magnetic energy therein.
 
An important point is that all models predict that the CME will contain an MFR after eruption. Indeed, MFR-like structures are often detected in CME EUV \citep[e.g.][]{dere1997,ZhangJ&al2012,vourlidas2014} and coronagraphic observations, \citep[e.g.,][]{vourlidas2013,vourlidas2017}
and in in-situ measurements \citep[e.g.,][]{burlaga1981,nieves2018}. Since the ejected structure is an MFR, white-light coronagraphic or in-situ observations should not generally be expected to provide a direct way to determine whether the pre-eruptive configuration was an SMA or an MFR. 

Although there seem to exist only two possible pre-eruptive magnetic geometries (SMA and MFR), several physical mechanisms (e.g., shearing, flux emergence, flux cancellation, helicity condensation) can give rise to either of them. There is a vast literature on the subject but there is no consensus on where and when given mechanisms may be relevant, if they need to operate alone or in tandem, and whether they are also the cause or the trigger of the subsequent eruption. Obtaining a clearer picture of what is the pre-eruptive configuration and how it forms will have important implications for the physical understanding of CMEs and the origin and evolution of CME-prolific active regions (ARs). Our improved understanding will help us in turn to better evaluate the eruptivity of a given AR and hence improve our predictive and forecasting abilities for space weather purposes \citep[e.g.,][]{vourlidas2019}. 

Despite significant advances in modeling, theory, and observational capabilities, the resolution of these issues is hampered by a number of factors: (i) limitations of observations (e.g., lack of routine magnetic-field observations above the photosphere, line-of-sight confusion in imaging), (ii) limitations of models (e.g., idealized boundary conditions, high numerical diffusion in the MHD codes), (iii) inconsistent application of terms and definitions (defining an MFR in the observations or the onset time of an eruption, for example) to observations leading to ambiguous conclusions, and perhaps most importantly (iv), MFRs and SMAs can be nearly indistinguishable in either field line plots or observations (e.g., filament threads).

Addressing these issues was the motivation behind the formation of an ISSI team tasked with ‘decoding the pre-eruptive configuration of CMEs’. The team met twice with three objectives: (1) debate the formation and configuration of filament channels, which comprise the observational manifestations of MFRs and SMAs; (2) identify the open \emph{and closed} issues; and (3) propose a path forward, for both modeling and observations, to resolve these issues. The results of our discussions form the core of this manuscript. We do not attempt to provide a comprehensive review of CME initiation and formation processes since there exist several recent excellent reviews on the matter \citep[][]{chen2011,cheng2017,manchester2017,green2018,archontis2019,georgoulis2019,liu2020}. We focus on three key questions: (1) what constitutes an MFR or SMA?; (2) how can MFRs/SMAs be identified in the solar atmosphere?; and (3) how do they form? We compile a set of recommendations for solving these outstanding  issues in the future, whether via modeling or improved observational data.

This paper is organized as follows. First, we provide definitions for the most important terms, such as MFR, SMA, Filament Channel, etc., to provide a clear baseline for the discussion. In Section~\ref{s:FC-formation}, we review the three dominant mechanisms for filament channel formation: flux emergence, flux cancellation, and helicity condensation. In Section~\ref{s:configurations}, we discuss the modeling expectations and observational signatures for 
SMAs and MFRs. In Section~\ref{s:recap}, we summarize our findings and provide a  table containing MFR/SMA observables that could be checked against observations. 
In Section\,\ref{s:path}, we present  recommendations for resolving the nature of the pre-eruptive configuration and its formation 
and in the Appendix~\ref{ss:path_approach} we demonstrate an appropriate methodology using MHD simulations and synthetic coronal images.

\subsection{Basic Definitions}\label{sss:defi}
This section contains definitions for the key terms 
discussed throughout the manuscript. The definitions are guided by practical considerations, i.e, to help interpret observations and models.
 \begin{figure*}[h]
\centering
\includegraphics[width=\textwidth]{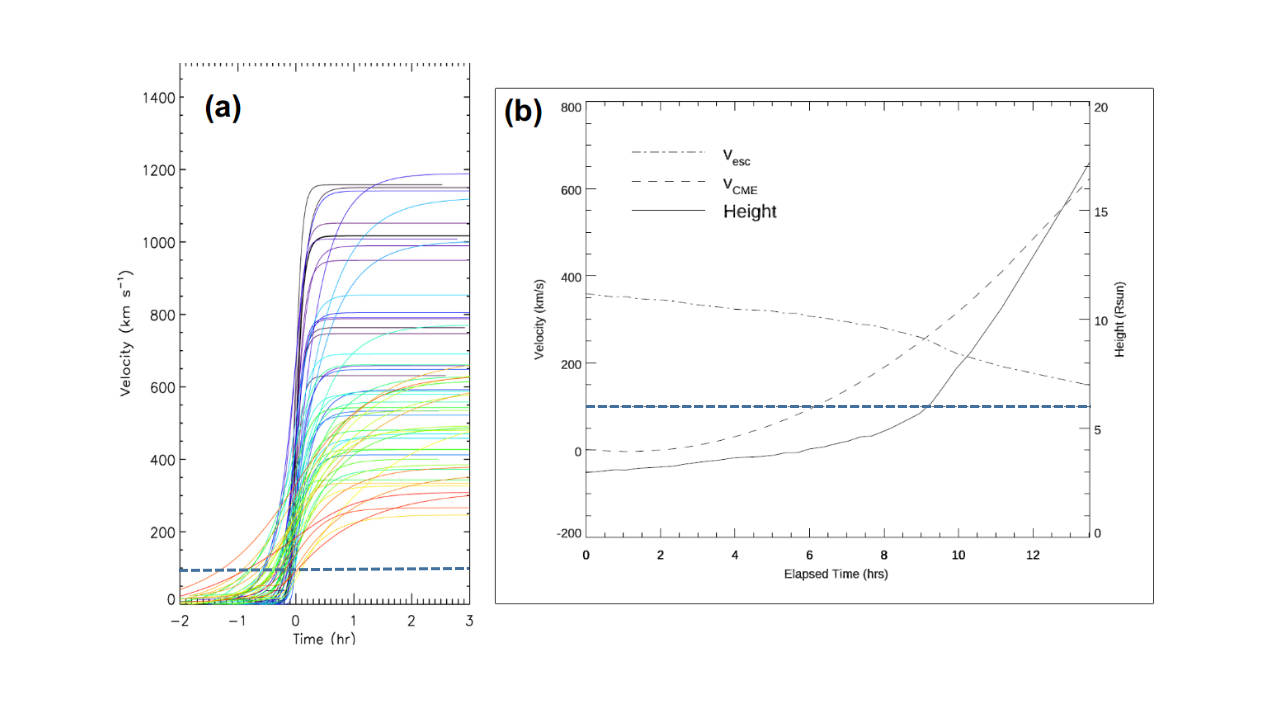}
\caption{Panel (a): superposed epoch analysis of the velocity-time profiles of 42 CMEs associated with eruptive flares. The key time of the profiles
(i.e., time=0) corresponds to the time of maximum acceleration for each CME. 
Modified  from \citet[][]{zhu2020} \copyright AAS. Reproduced with permission. Panel (b): velocity-time profile of a streamer blowout CME (dashes).
Adapted from \citet[][]{vourlidas2002}. 
In both panels, the horizontal dashed blue lines correspond
to a speed of 100 km/s}
\label{fig000}
\end{figure*}

\textbf{Pre-Eruptive Condition:} This is the key term for our discussion, yet it is hard to define precisely. Eruptions occur over a wide range of time scales and exhibit a variety of signatures at the surface and in the lower atmosphere. For example, some eruptions precede, others follow, the soft X-ray (SXR) flare onset 
and some are not accompanied by flares at all. The timing of, or rather the determination of whether an eruption is under way, is crucial in assessing the physical processes responsible for it (ideal or non-ideal, for example). Since there is no widely-accepted measure or definition of the start of an eruption, we put forth one as follows. It is logical to expect that a CME will occur when the speed of the rising structure exceeds a considerable fraction of the local Alfv\'en speed. At that point, the rising structure can no longer be considered in a state of a quasi-static rise caused by, for instance, slow photospheric motions. This quantity is, however, very difficult to assess observationally or via simulations since the coronal magnetic environment is insufficiently known. Instead, we take a more practical, and conservative, approach and consider a high enough speed to ensure that a CME will occur across the range of CME source regions. We posit that 100 km s$^{-1}$ is a practical limit based on observations of both flare-related and streamer-blowout CMEs (e.g., Figure~\ref{fig000}).  
Hence, we propose the following definition for the pre-eruptive condition: \textit{The pre-eruption phase ends when the speed of the rising (and eventually, released) magnetic structure exceeds 100 km s$^{-1}$.}\\

\begin{figure*}[h]
\centering
\includegraphics[width=\textwidth]{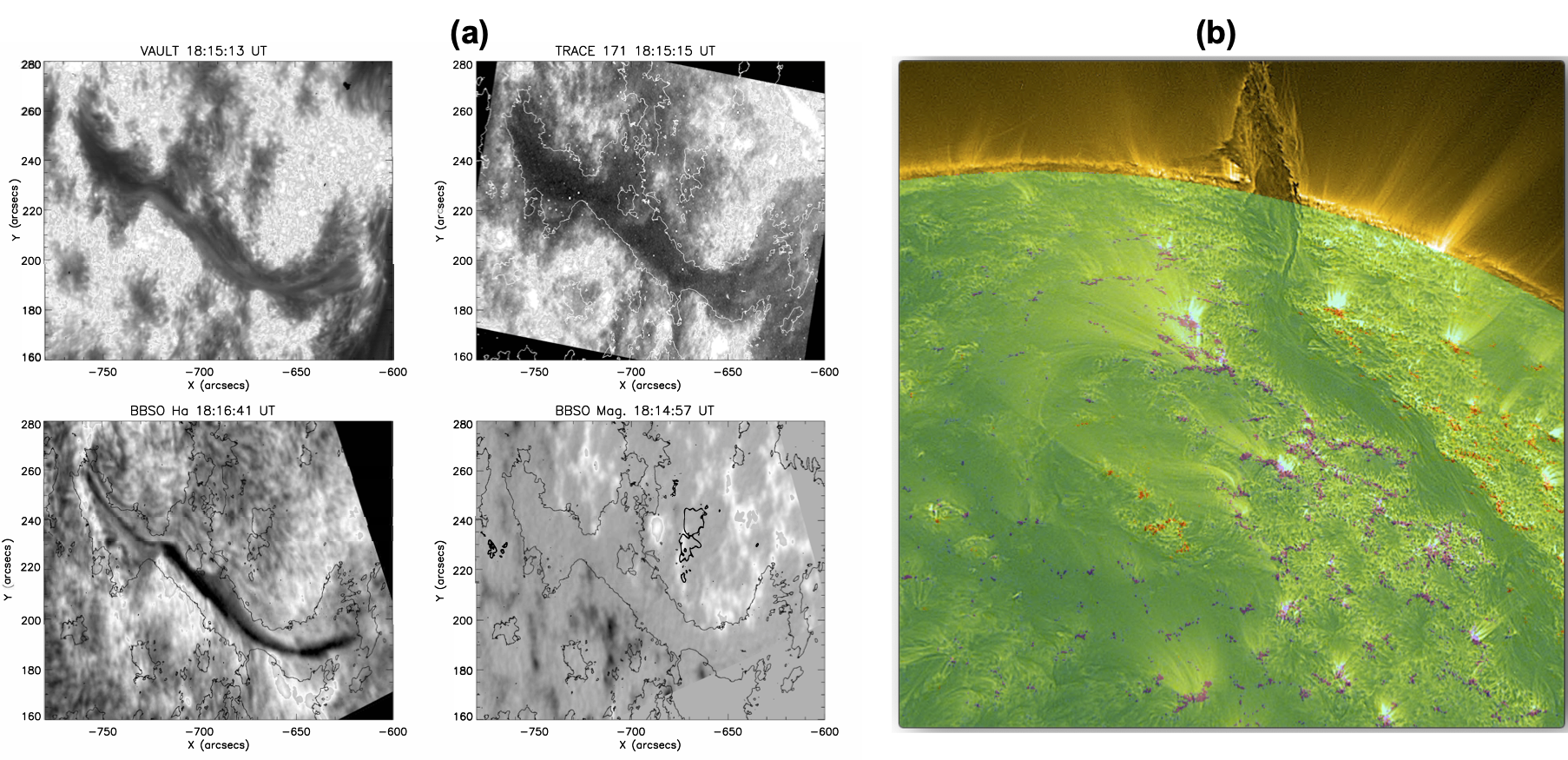}
\caption{Panel (a): A filament channel with prominence material observed in Lyman-$\alpha$ (top left), 171\AA\ (top right), H$\alpha$ (bottom left) and photospheric magnetic field (bottom right). The contours delineate the extent of the Lyman-$\alpha$ prominence. From \citet[][]{vourlidas2010}. \textit{\copyright 2009, Springer Nature}. Panel (b): Composite of AIA 171\AA\ (gold) and HMI magnetogram (green overlay with positive flux in red and negative flux in blue) of a spectacular filament channel containing a filament/prominence. 
}
\label{fig00}
\end{figure*}

\begin{figure*}[h]
\centering
\includegraphics[width=0.99\textwidth]{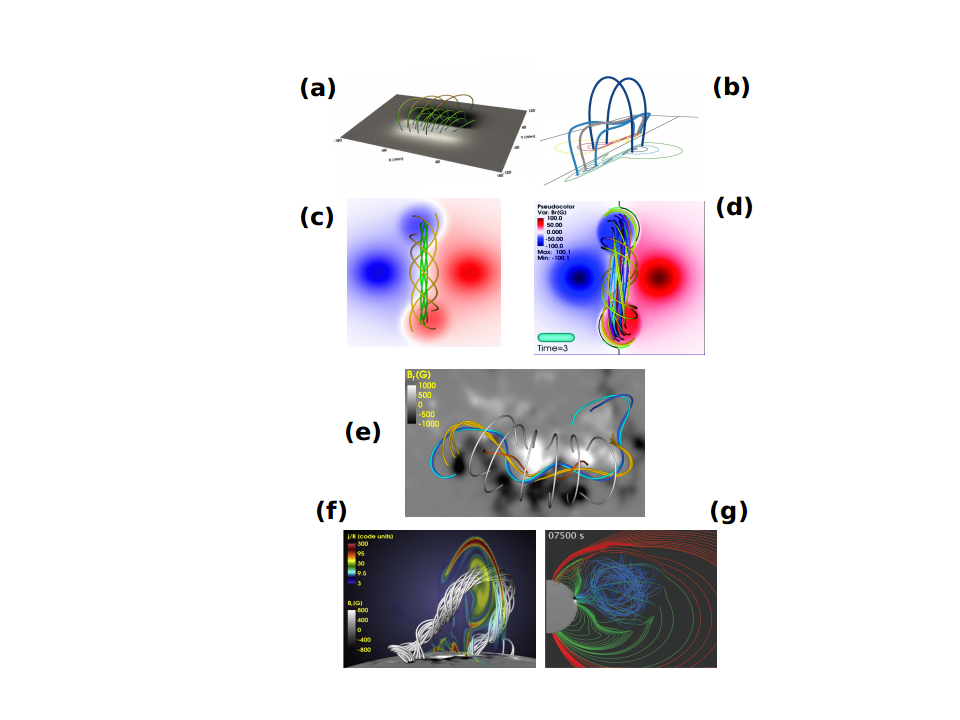}
\caption{Pre-eruptive (panels a-e) and eruptive configurations (panels f-g) from
MHD simulations. Magnetic fields lines are shown in all panels.
Panel (a): 2.5D SMA. Modified from \citet[][]{zhou2018}. Panel (b): 3D SMA. Modified from \citet[][]{devore2000p}. Panel (c): MFR with twisted field lines  concentrated in the core of the configuration. Modified from \citet[][]{titov14}. Panel (d): MFR with twisted field lines concentrated in the periphery of the configuration (i.e., hollow-core configuration). Modified from \citet[][]{titov14}. Panel (e): hybrid configuration.  Modified from \citet[][]{torok18b}. Panel (f): eruptive MFR from a simulation employing a pre-eruptive MFR that has evolved from the [hybrid] configuration shown in panel (e)
. Modified from \citet[][]{torok18b}. Panel (g): eruptive MFR from a simulation employing a pre-eruptive SMA. Modified  from \citet[][]{2008ApJ...683.1192L}. All figures are \copyright AAS and reproduced with permission.
} \label{fig01}
\end{figure*}

\noindent\textbf{Polarity Inversion Line (PIL):}  The line separating areas of opposite magnetic polarity on the Sun. 

\noindent\textbf{Filament channel (FC):} 
Filament channels are the upper atmospheric counterparts of PILs. They correspond to regions where the magnetic field is largely aligned with the photospheric PIL. FCs are identified in the chromosphere by the orientation of chromospheric fibrils. When partially filled with cool and dense plasma, FCs manifest themselves as filaments or prominences in chromospheric and coronal observations e.g, see Figure \ref{fig00}.\\

\noindent\textbf{Sheared Magnetic Arcade (SMA):} A set of field lines (or flux bundles) that cross a PIL with an orientation deviating from the local normal to the PIL (see Section\,\ref{ss:SMA}). An SMA does not have an axis field line about which its flux, or an inner part of its flux, twists from end to end but it can nevertheless contain a small amount of twisted flux. In a strongly sheared arcade, the orientation of the field lines closely follows the PIL. \textit{In an SMA, the sheared flux is much larger than the twisted flux\/}. SMA models range from simple, pseudo-2.5D cases that contain only arched field lines of the same orientation (Figure \ref{fig01}(a)) 
to more complex, fully 3D, cases that contain S-shaped and dipped field lines (Figure \ref{fig01}(b)). In this review, we  mainly discuss these more realistic, complex SMAs.\\

\noindent\textbf{Magnetic Flux Rope (MFR):} A twisted flux tube where the majority of the interior field lines wind about a common axial field line along the length of the tube. This is an expanded definition of the textbook MFR \citep{Priest2014} to account for the complexities encountered in current models and their comparisons with observations, which we discuss in detail in the rest of the paper. MFRs are characterized by the presence of a magnetic axis, a current channel, and twist extending 
over the full length of the magnetic axis (Figure \ref{fig01}(c)). 
We classify MFRs based on the twist number $N$ (their end-to-end number of turns) as:
1) weakly twisted ($N<1$);
2) moderately twisted ( $N \approx (1-2)$);
3) highly twisted ($N>2$).
MFRs do not necessarily possess uniform twist, or twist peaking in the vicinity of the axis (Figure 2(c))
--- field lines near the axis can have minimal twist (Figure 2(d)).
However the defining MFR characteristic is that \textit{the twisted flux is much larger than the sheared flux\/.} The axis of MFRs relevant to eruptions follows the PIL relatively closely. For an arched MFR to possess dips, one must additionally require at least $\sim 1$ end-to-end turn (see Section\,\ref{ss:MFR} for details).
Examples of eruptive MFRs resulting from pre-eruptive MFRs
or SMAs are given in panels (f) and (g) of Figure~\ref{fig01} respectively.
\\

\noindent\textbf{SMA-MFR Hybrid:} 
A magnetic configuration that contains both sheared and twisted flux in non-negligible, and possibly evolving, proportions (Figure~\ref{fig01}(e)). Hybrids can arise in FCs by magnetic reconnection in the center of an SMA (Section~\ref{s:FC-formation}) or between an SMA and its overlying potential loops  or by the diversion of some of the current-carrying flux to localized polarities along the PIL (Section~\ref{ss:SMA}).\\

All three magnetic field configurations above are current-carrying structures loaded with free magnetic energy and magnetic helicity. Hence, they could all lead, under specific conditions, to eruptions. In the rest of the paper, we discuss these conditions, caveats and observational and modelling challenges.

\section{\textbf{Filament Channel} Formation}
\label{s:FC-formation}
Filament channels are the cornerstones for understanding solar eruptive activity. FCs are \textit{sheared}, and hence are reservoirs of magnetic free energy, which is required to fuel eruptive phenomena. 
CMEs occur only above PILs that are traced by an FC
(all CME theories and models require the presence of an FC above a PIL). Hence, understanding FC formation equates to understanding the pre-eruptive configuration in the corona. FC properties are reviewed in several papers \citep[e.g.,][]{martin1998,mackay2015}. 

The literature abounds with modeling efforts to explain the formation of the magnetic configuration of a filament channel  \citep[for a comprehensive review, see][]{mackay2010}. 
The models can be differentiated between those employing surface effects (differential rotation, meridional flows, shear and converging flows, helicity condensation), and those employing sub-surface effects (the emergence of MFRs). In particular, FC could be formed by the following
three mechanisms: (a) flux emergence, (b) flux cancellation, and (c) helicity condensation. Note, that frequently these mechanisms do not operate in isolation. Sections \ref{ss:emergence}, \ref{ss:cancellation}
and \ref{ss:condensation} discuss these mechanisms and 
the corresponding observations in detail.

\subsection{Flux Emergence}
\label{ss:emergence}

\subsubsection{Introduction}   
A number of observational studies have shown that  the formation of FCs is related, implicitly or explicitly, to the process of magnetic flux emergence from the solar interior into the solar atmosphere.
There are studies suggesting that during flux emergence, a (twisted) flux tube, which rises from the solar interior, can emerge as a whole above the photosphere 
\citep[e.g.][]{lites2005,okamoto2008,lites2010,xu2012,kuckein2012}. 
This bodily emergence may result in a pre-eruptive MFR configuration. 
Other studies have reported that pre-eruptive structures can form 
along the strong PILs of ARs during flux emergence. For instance, the gradual build-up of free energy injected by photospheric motions can lead to shearing of the emerged magnetic field, forming an SMA. 
Numerical simulations of magnetic flux emergence indicate that such an SMA may eventually evolve into an MFR, which could erupt towards the outer solar atmosphere \citep[e.g.][]{manchester2004,archontis08a, Archontis_etal2009,Fan_2009,archontis2012,Moreno-Insertis_etal2013,leake13a, Leake_etal2014,Syntelis_etal2017,Syntelis_2019b}. 
Moreover, studies have reported the formation of FCs at the periphery of ARs or between neighbouring ARs \citep[e.g.][]{gaizauskas1997,wang2007}.

Here, we should highlight that the scope of this review is not an extended presentation of the various aspects of magnetic flux emergence, as a key process towards understanding the nature of solar magnetic activity. There exist several 
comprehensive reviews on this subject \citep[e.g.][]{moreno-insertis07,2008JGRA..11303S04A,fan2009,nordlund2009,archontis12,hood2012,stein2012,cheung2014,toriumi14,archontis2019,leenarts2020}. 
Thus, in the following sections, we mainly focus 
on the pre-eruptive structures  that form as a result of
flux emergence, from a modelling (section 2.1.2) and an observational perspective (section 2.1.3).

\subsubsection{Flux Emergence Modeling}

Numerical experiments of magnetic flux emergence can be classified in two very broad categories, based on the nature of the background atmosphere that is included in the simulations.
In the first category, the numerical domain contains the upper part of the solar atmosphere (e.g., the solar corona), and the magnetic flux is injected through the lower boundary of the domain, typically in the form of an MFR \citep[e.g.][]{Fan_etal2004}. Such boundary-driven models are useful to study 
the stability and/or the eruptive behavior of coronal structures, but they do not capture the emergence process itself.
In the second category, the magnetic field emerges from the solar interior and expands into a highly stratified atmosphere above the photosphere, driving the atmospheric dynamics self-consistently \citep[e.g.][]{Manchester_2001, Fan_2001, Archontis_etal2004}.
A typical (idealized) numerical setup for the experiments in the second category consists of a convectively stable solar interior, a photospheric/chromospheric region of constant temperature and decreasing density, a region where the temperature increases rapidly with height, mimicking the temperature gradient of the transition region, and an isothermal corona. In this review, we will mainly focus on the models of the second category, as they are more suitable for studying the formation and evolution of pre-eruptive magnetic structures, at various atmospheric heights.

The most common initial condition for the sub-photospheric magnetic field is 
a twisted flux tube  placed in
the upper part of the solar interior, adopting the configuration of a straight horizontal tube or of a torus-shaped tube.
Then, the emergence is initiated by imposing a density deficit along the tube, which makes part of the tube magnetically buoyant \citep[e.g.][]{Fan_2001}, or by imposing a velocity perturbation \citep[e.g.][]{Magara_etal2001, Magara_etal2003} along a segment of the tube, which leads to the development of a rising loop with an $\mathrm{\Omega}$-like shape \citep[e.g.][]{Archontis_etal2004,manchester2004}.
Toroidal flux tubes are typically used to mimic the top part of a sub-photospheric $\mathrm{\Omega}$-shaped emerging loop \citep[e.g.][]{Hood_etal2009,Cheung_etal2010}.

A considerable number of flux-emergence simulations have been using a sub-photospheric horizontal magnetic flux sheet as initial condition. The interplay between the effect of convective motions on the magnetic field and the effect of the distorted field on the motion leads to the development of a series of small-scale interconnected $\mathrm{\Omega}$-shaped loops \citep[a `sea-serpent' configuration][]{pariat2004}) that may eventually emerge through the photosphere \citep[e.g.][]{Isobe_etal2007, Archontis_etal2009b, 
Toriumi_etal2010,Stein_etal2011, Stein_etal2012}. 
These simulations have been used successfully to study small-scale dynamic phenomena such as Ellerman bombs and UV bursts \citep[e.g.,][]{Danilovic_etal2017,Hansteen_etal2017,Hansteen_etal2019} and the formation of complex bipolar regions and pores \citep[e.g.][]{Stein_etal2012}. However,
these numerical experiments have not (yet) been able to produce  
large-scale pre-eruptive configurations or eruptions.

We note that a significant number of flux-emergence simulations  incorporated additional physics, such as convective motions, radiative heating and cooling, heat conduction, ambipolar diffusion, ion-neutral interactions, and non-equilibrium ionization \citep[e.g.,][]{Leake_etal2006,Stein_etal2006,Cameron2007,MartinezSykora2008,Isobe_etal2008,Cheung_etal2010,Fang_etal2010,Rempel2014,Chen2017,Hansteen_etal2017,MorenoInsertis2018,NobregaSiverio2018,cheung2019,Toriumi2019}. 
These simulations are necessary for studying the thermodynamical aspects of phenomena related to flux emergence and the atmospheric response to the dynamic emergence of solar magnetic fields.
However, the vast majority of these experiments have not yet included the solar corona to an extent that is required to model pre-eruptive configurations. 

In the following, we will review the formation and evolution of the most common pre-eruptive configurations that
are found in flux-emergence simulations.

\begin{center}
\textbf{Partial and bodily emergence of MFRs}    
\end{center}

Numerical simulations have shown that the rise of an initially horizontal sub-photospheric flux tube is usually not followed by the bodily emergence of the tube into the higher atmosphere \citep[e.g.,][]{Fan_2001,Magara_etal2003,manchester2004}. When the top part of a flux tube starts to emerge into the atmosphere, it expands rapidly due to the pressure difference between itself and the solar atmosphere. During this expansion, a region with low density and pressure is formed at the centre of the emerging region, at photospheric heights. Plasma then drains along the field lines, moving from the top part of the flux tube towards this low-pressure region \citep[e.g.][]{manchester2004} and, eventually, it accumulates in the dips of the twisted field lines. The dips become heavier and, thus, the axis of the tube cannot emerge fully into the corona. Rather, it reaches only a few pressure scale heights above the photosphere. Full (bodily) emergence can occur only if the tube is buoyant enough to reach the top of the low pressure region, before the drained plasma accumulates at its dips. 

\citet{Magara_etal2003} demonstrated how the curvature of different field lines within a horizontal flux tube, which emerged from just below the photosphere (-2.1~Mm), affected the draining of plasma. \citet{Murray_etal2006} performed a parametric study in a similar setup, and varied the initial twist (affecting plasma draining) and the initial magnetic field strength (affecting the buoyancy) of the sub-photospheric flux tube. More draining and more buoyancy assisted the axis of the tube to move higher inside the photosphere. \citet[][]{Mactaggart_2009} studied a case where the middle part of the tube was emerging and the flanks were submerging. This led to more efficient draining along the flanks of the tube but it didn't trigger bodily emergence. 

In all the above-mentioned studies, the initial location of the emerging flux tube was near the photosphere. \citet[][]{Syntelis_2019a} imposed horizontal flux tubes deeper in the solar interior (-18 Mm), to allow the tubes to develop a more strongly curved shape as they emerge towards the photosphere. They performed a large parametric study, varying the magnetic field strength, radius, twist, and length of the buoyant segment of the tube (a proxy for the curvature). They found that the axis of the flux tube always remained below the photosphere. In addition, they showed that it is non-trivial to predict the combined effects of these parameters during the emergence of the field. 
For instance, a non-intuitive result was that high-strength (weak-strength) flux tubes may fail (succeed) to emerge into the atmosphere, depending on their geometrical properties.

\begin{figure*}[h]
\centering
\includegraphics[width=0.5\textwidth]{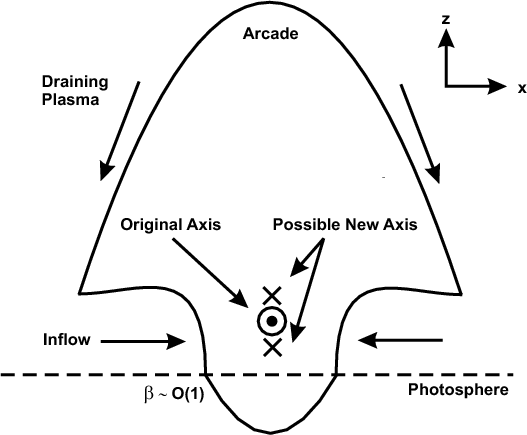}
\caption{Creation of converging flows during the partial emergence of a sub-surface MFR and subsequent reorganization of the coronal field, which produces a new MFR. Reproduced with permission from A\&A, \copyright  ESO. From \citet[][]{Mactaggart_2009}.} 
\label{fig2}
\end{figure*}

On the other hand, \cite{Hood_etal2009} studied the emergence of a weakly twisted, toroidal flux tube. They found that the geometrical shape of the tube (higher curvature), and the smaller number of dips along the twisted field lines, triggered sufficient plasma draining and buoyancy, which helped the tube to emerge bodily into the corona, given that the initial field strength in the tube was chosen sufficiently strong. \citet{Mactaggart_2009} systematically varied the field strength and found cases where the axis of the tube (i) could not break through the photosphere, (ii) emerged but stayed within the photospheric layer, and (iii) reached coronal heights and continued rising (i.e., emerged bodily). The bodily emerged MFRs in these simulations appear to be weakly twisted (about one turn or less), but the dependence of the MFR twist on the initial field strength or twist of the sub-photospheric flux tube has not yet been studied systematically. Both \cite{Hood_etal2009} and \citet{Mactaggart_2009} reported that a second MFR formed via magnetic reconnection (see below for a detailed description of this process) underneath the bodily emerged MFR. If, on the other hand, the original axis did not emerge into the corona, the second MFR was seen to form above it (see Figure \ref{fig2}).

The numerical experiments summarized in this section suggest that whether a sub-photospheric twisted flux tube emerges bodily, partially, or not at all, depends mainly on the properties of the rising magnetic field and its geometric configuration. It appears that bodily emergence requires rather specific conditions, namely a toroidal geometry and a relatively large field strength. We discuss observational evidence for bodily emergence in Section \ref{sss:fe_obs}. Next, we focus on the formation of MFRs via magnetic reconnection, as seen in flux-emergence simulations.  

\begin{center}
\textbf{MFR formation via  reconnection}
\end{center}

The emergence of a single sub-photospheric flux tube typically
leads to the formation of a bipolar region, whose polarities separate over time. 
Photospheric motions commonly found in the vicinity of the region's PIL include: (i) shearing along the PIL, resulting from the Lorentz force developed at the photosphere due to the expansion of the emerging field \citep[e.g.,][]{Fan_2001,Manchester_2001}, (ii) rotation of the polarities driven by the propagation of a torsional Alfv\'{e}n wave into the atmosphere \citep[e.g.,][]{Longcope_etal19997,Fan_2009,leake13a,Sturrock_etal2015,Sturrock_etal2016}, and (iii) downflow of plasma towards the low-pressure region above the PIL as discussed above \citep[e.g.,][]{manchester2004}. The combination of these motions induces converging flows towards the PIL \citep[e.g.,][]{Archontis_2010,Syntelis_etal2017}, which ultimately lead to the formation of an MFR.

\begin{figure*}[h]
\centering
\includegraphics[width=0.7\textwidth]{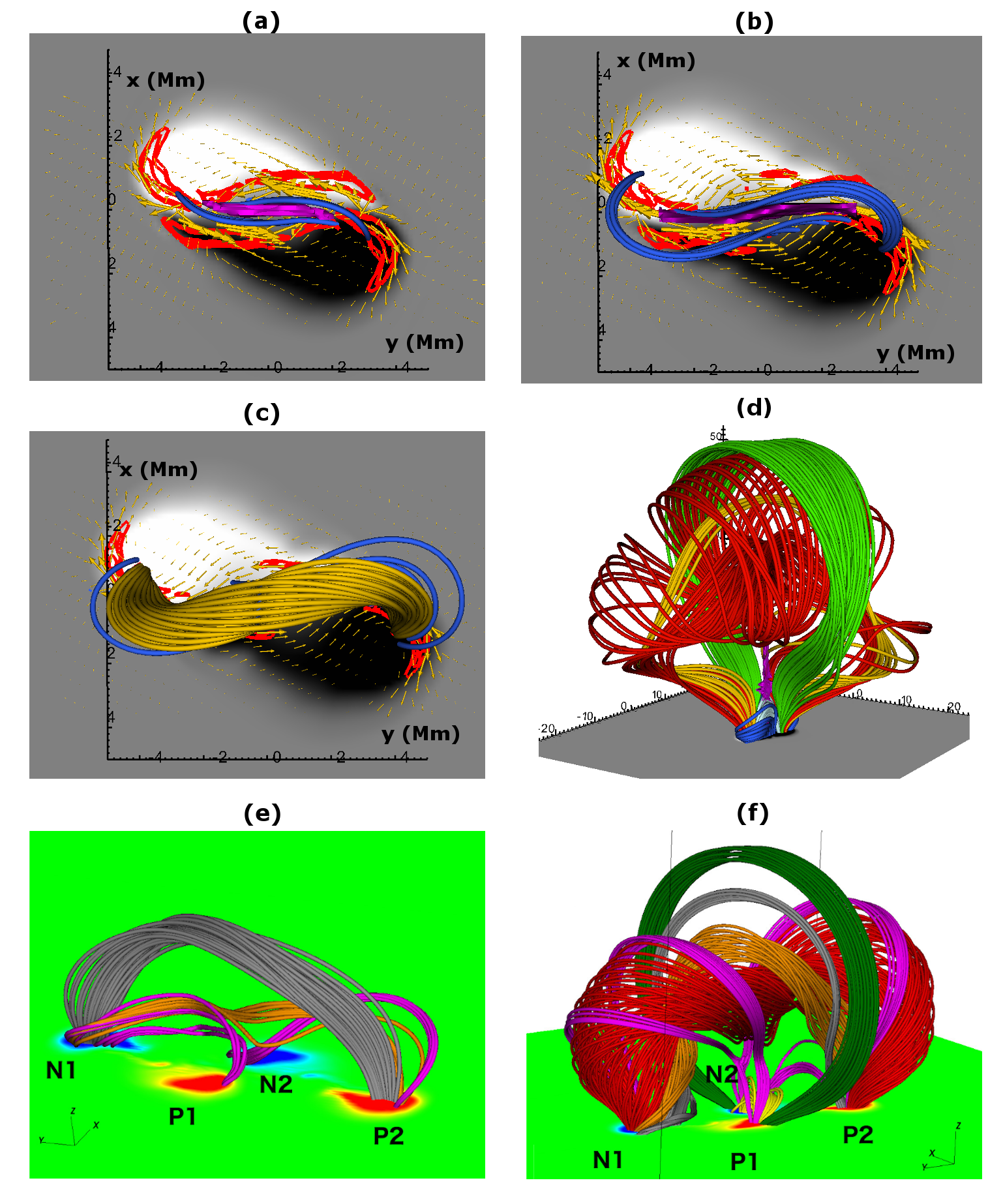}
\caption{MFR formation by reconnection in bipolar and
quadropolar regions. 
    (a, b) Top view on field lines that become increasingly sheared and adopt a double-J-like shape (blue lines). The horizontal slice shows the distribution of $B_{z}$ (black and white) at the photosphere. Over-plotted are yellow arrows showing the photospheric velocity field scaled by magnitude, red contours showing the photospheric vorticity and a purple isosurface showing $|\mathbf{J}|/|\mathbf{B}|=0.3$
    (c) Field lines of the sigmoidal MFR formed by reconnected J-like field lines (orange). 
    (d) The field line topology during the eruption of the MFR. The red lines show tether-cut field lines due to reconnection, the green lines show the remaining envelope field which has not yet reconnected and the cyan lines show the post-reconnection arcade. The purple isosurface shows the flare CS. (e) The topology of the field lines in a quadrupolar region. Pink lines show the field lines of the two interacting magnetic lobes. Orange lines have been traced from the close vicinity of the core of the MFR. Field lines in grey color show the "envelope" field resulting from reconnection between the two magnetic lobes.
(f) The field line topology during the rise of the MFR, which is eventually confined by the envelope field. The red lines are field lines, which have been reconnected during the interaction of the two magnetic lobes and the formation and rise of the MFR. The green and grey field lines lines belong to the envelope field above the MFR. The low-lying yellow field lines are reconnected field lines, which form a system of post-reconnection arcade loops that connect the inner polarities (P1, N2).(from \citet[][]{Syntelis_etal2017} and \citet[][]{Syntelis_2019c} ). 
    \copyright AAS. Reproduced with permission. 
}
\label{fig1}
\end{figure*}

This is illustrated in Figure~\ref{fig1}. The described motions inject free magnetic energy into the system and lead to the gradual formation of an SMA above the PIL. Eventually, a strong, vertical current layer is formed within the SMA (Figure~\ref{fig1}a), the sheared field lines adopt a double-J shape (Figure~\ref{fig1}b), and the current layer's strength increases. Field lines that belong to the SMA reconnect above the PIL and form a (new) MFR with a sigmoidal shape (Figure~\ref{fig1}c) at low atmospheric heights. Spectroscopic observations by CDS,EIS and IRIS, taken at the footprints of forming MFRs, showed significant line-shifts and line-broadenings, implying the occurrence of reconnection in the low atmosphere \citep[][]{foley2001,harra2013,cheng2015}.

The newly formed MFRs are typically weakly twisted (approximately one turn, although the exact number of turns has not been measured in most studies) and can become eruptive (Figure~\ref{fig1}d). Overall, a large number of numerical experiments \citep[e.g.][]{manchester2004,archontis08a, Archontis_etal2009,Fan_2009,archontis2012, 
leake13a,Moreno-Insertis_etal2013,Archontis_etal2014,Fang_etal2014,Leake_etal2014,Lee_etal2015,Syntelis_etal2017,Toriumi_etal2017,Syntelis_2019b,Syntelis_2019c} have shown that post-emergence MFRs can form through reconnection of sheared magnetic field lines across a current layer above the PIL of an emerging AR, in the same manner as suggested for the formation of large-scale (quiescent) FCs suggested by \citet[][]{vanballegooijen89}; see also Section~\ref{ss:cancellation}.

Flux-emergence models have been used to study also the dynamic evolution 
at the PILs between colliding/interacting bipoles in quadrupolar ARs \citep[e.g.][]{Fang_2015, Takasao_2015,Toriumi_etal2017,cheung2019,Syntelis_2019c}. We note that the nature of shear and its role for the formation and eruption of MFRs has been investigated also in observational studies, for cases of two colliding emerging bipoles \citep{chintzoglou2019}. Here we focus on the formation of 
MFRs in quadrupolar ARs in flux-emergence simulations.

The most common initial set up of the numerical simulations on this subject include the emergence of either:
(i) a single $\mathrm{\Omega}$-loop flux tube with low twist \citep[e.g.][]{Murray_etal2006, Archontis_etal2013, Syntelis_etal2015};
(ii) a single flux tube emerging at two different locations along its length (``double-$\mathrm{\Omega}$'' loop), so that two nearby bipoles appear at the photosphere \citep[e.g.][]{Fang_2015,Lee_etal2015,Toriumi_etal2017};
(iii) two different $\mathrm{\Omega}$-loop flux tubes emerging nearby \citep[e.g.][]{x2.2Toriumi, Toriumi_etal2017};
(iv) a single kink unstable flux tube (such flux tubes, depending on the twist, can form from bipolar to more complex multipolar configurations) \citep[e.g.][]{Takasao_2015,Toriumi_etal2017,Knizhnik_2018}.
In the above cases, a PIL is formed between the inner-most polarities of the quadrupole, above which a strong current layer can build up and a MFR can be formed. Such MFRs can potentially erupt.

\citet{Syntelis_2019c} reported on a model of recurrent confined eruptions in a quadrupolar region. Initially, prior to the formation of the MFR, the two magnetic lobes above the two emerged bipoles were not interacting. As the bipoles emerged and moved, the inner polarities of the quadrupole approached each other, and a current sheet (CS) formed above the PIL between them, extending between the two magnetic lobes. The strength and size of this CS progressively increased over time. Reconnection between the two magnetic lobes through that CS formed a magnetic ``envelope'' above the quadrupolar region and a weakly-twisted (approximately one turn) post-emergence low-lying MFR (Figure~\ref{fig1}e). This pre-eruptive MFR did not result from the internal reconnection of a SMA. Rather, the MFR formed directly. The eruption of the MFR was triggered by reconnection internally within the quadrupolar region, occurring both below and above the MFR. During the confined eruption, the weakly twisted low-lying MFR became a larger and more twisted confined coronal MFR, similar to confined-flare-to-flux-rope observations \citep[e.g.][]{Patsourakos&al2013}. The eruptivity of the latter MFR was not studied.

Post-emergence MFRs can form in a recurrent manner as long as the photospheric motions are present and magnetic energy is injected into the system \citep[e.g.][]{Moreno-Insertis_etal2013, Archontis_etal2014, Syntelis_etal2017,Syntelis_2019b,Syntelis_2019c}. When the driving motions are no longer able to build enough free energy, the recurrent formation of post-emergence MFRs ceases. It is important to note here that in flux-emergence models, the formation of MFRs and subsequent eruptions do not occur only during the flux-emergence phase (i.e., while the photospheric flux still increases). They can also occur when the photospheric flux has saturated, but the photospheric motions are still present
(e.g, the vertical dashed lines in Figure \ref{fig14} in the Appendix).

As we have mentioned previously in this section, the pre-eruptive MFRs are typically weakly twisted. However, we should highlight that the number of turns of the field lines resulting from the reconnection between two flux systems depends on the specifics of the reconnection region. For instance, \citet{Wright_2019} showed that: a) the relative orientation of the footpoints of the two pre-reconnection flux systems and b) their magnetic helicity content
(for a discussion on magnetic helicity see Section \ref{ss:condensation}), are crucial to determine the resulting twist after they reconnect to each other. These two factors affect how self-helicity is partitioned between the resulting post-reconnection systems. \citet{Wright_2019} discussed cases where the twist of the resulting two flux systems can increase, decrease or remain the same after their interaction.
\citet{Priest_2020} further studied how self-helicity is partitioned between reconnecting systems, by examining twist during the reconnection of flux sheets, sheaths and tubes, and discussed how the twist of the erupting MFR increases during the eruption, leaving behind an untwisted arcade. Multiple reconnection events between different flux systems can furthermore increase/decrease the twist of a pre-eruptive MFR. This can occur when the post-reconnected field lines reconnect again with field from the same pre-reconnected systems, or when they reconnect with other flux systems. Similarly, during an eruption, the MFR twist typically increases due to multiple reconnection events in the flare CS below the core of the erupting MFR (e.g., red lines in Fig.~\ref{fig1}d and  \citet[][]{gibson06, Syntelis_etal2017, Inoue_2018,Syntelis_2019b}).

\begin{figure*}[t]
\centering
\includegraphics[width=\textwidth]{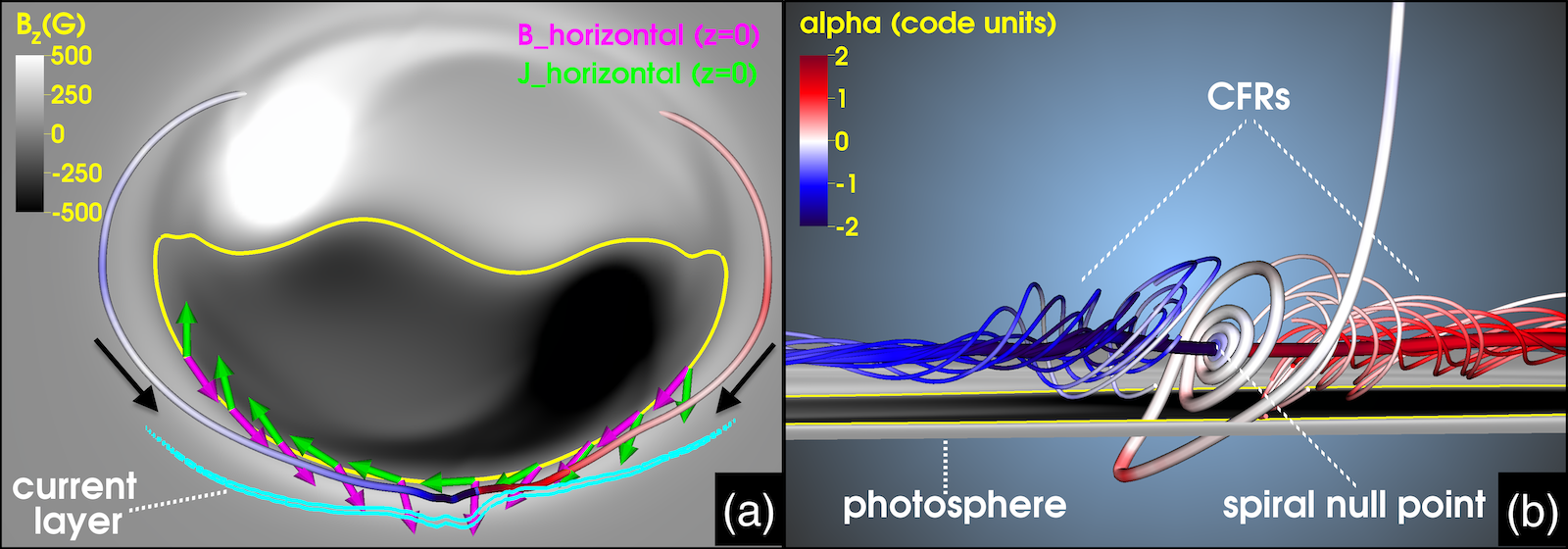}
\caption{
    MHD simulation of FC formation at the periphery of an emerging AR. 
    (a) Top view on the AR, which emerges into a positive background field, close to a pre-existing AR (not shown). The yellow line marks the PIL, arrows show the horizontal magnetic field (magenta) and current density (green) along the external PIL section. The magnetic field switches sign along and above this section (indicated by black arrows), leading to the formation of CFRs (i.e., two adjacent MFRs of opposite helicity sign) via reconnection across the current layer (indicated in cyan). 
    (b) Oblique view on the MFRs, colored by $\alpha=({\bf j} \cdot {\bf B})/B^2$. The location of the spiral null point between the two MFRs is illustrated by a thick field line.
}
\label{fig1a}
\end{figure*}

MFRs could also form at the periphery (or in between) ARs as was recently suggested by 
\citet[][]{torok18a}.
They modeled the emergence of an MFR close to a pre-existing bipolar AR. The orientation of the MFR was chosen such that a quadrupolar configuration with a current layer between the pre-existing and newly emerged flux systems resulted, similar to the configurations just described. A pair of so-called ``conjoined flux ropes'' \citep[CFRs; e.g.,][]{wyper14a,wyper14b,titov17,fu17} was created by the tearing instability in the current layer and advected to the lower atmosphere, leading to the formation of two MFRs of opposite axial-field direction (i.e., opposite helicity sign) that are located end-to-end above the ``external'' PIL section between the new and pre-existing flux (Figure\,\ref{fig1a}). Note that the CFRs form in addition to the main MFR that bodily emerges or forms  above the ``internal'' PIL of the emerging region, as described above. Note also that the presence of a pre-existing bipole is not required for this mechanism to work, as the current layer forms between the inner polarities. That is, the mechanism can work also when an AR emerges within, or close to, a single-polarity region, i.e., a coronal hole.

\subsubsection{Flux emergence observations}
\label{sss:fe_obs}
We start with observations of FC formation within the cores of emerging flux regions interpreted in favor of bodily emerging MFRs.
\citet{lites1995} studied the emergence and evolution of a small $\delta$-spot AR, where a low-lying filament was seen in H$\alpha$ above the PIL. They found that a ``magnetically closed structure'' remained in the corona even after the disappearance of the $\delta$-spot AR. They attributed this to an MFR in equilibrium with the ambient coronal field, after it had emerged through the photosphere. Later works revisited the emerging MFR scenario by examining the vector magnetic field in locations where filaments form in plages, at the photosphere \citep[][]{lites2005,okamoto2008,lites2010} and by also including chromospheric vector magnetograms \citep[][]{kuckein2012,xu2012}. The photospheric vector field in these works is found to be of a ``concave-up'' geometry suggesting a magnetic structure dipping at the PIL, and also that of an ``inverse'' horizontal field (with respect to that of a potential field, which is called ``normal'') crossing the PIL. To our knowledge, these works are the only references suggesting full (i.e., bodily)  MFR emergence, making it a rather scarce observation. 

However, several important observational limitations (e.g., incomplete spectral and temporal coverage or confusion from pre-existing structures and the lack of multi-height vector magnetic field measurements),
do not allow to  unambiguously interpret observations of flux emergence. Therefore, their interpretation in terms
of bodily emerging MFRs could not be unique. 

For example, several analyses take place while a filament already exists in these locations \citep[][]{okamoto2008,kuckein2012}, casting doubt that these are studies of true FC formation. In addition,
MHD modeling of dynamic emergence of a twisted cylinder into an overlying arcade by \citet[][]{vargas2012} showed the same photospheric signatures as observed by \citet[][]{okamoto2008}. However, this was not a case of a bodily-emerging flux rope into the corona,
since the axis of the emerging cylinder reached only one photospheric scale-height above
the photosphere.
This result reveals that the concave upward geometry is not an exclusive indication of a bodily emerging MFR. The \citet[][]{xu2012} study was based on the analysis of a single chromospheric magnetogram (in HeI 10830 \AA) and comparison to H$\alpha$.
They found that the chromospheric magnetic field exhibited the normal
configuration while the photospheric magnetic field
was concave upwards and these results 
were interpreted in terms of  
bodily emergence of a flux rope that is producing a filament. The authors acknowledge that their interpretation is not unambiguous given that the observed differences in HeI 10830 \AA \, and H$\alpha$ may be  
interpreted in terms of optical-depth
effects. The same issue affects the \citet[][]{kuckein2012} results since they used the same lines. 
\citet[][]{mackay2010} analyzed the same sequence as \citet[][]{okamoto2008} but interpret the signatures in the Ca II H line 
being due to a rising SMA rather than an MFR. \citet[][]{lites2010} discuss the possibility of surface flows creating their observed short-lived filament channel via cancellation, but they reject it based on the lack of systematic photospheric flows that could  produce significant shear and convergence along the  channel. However, their flow field was determined from the photospheric granulation instead of tracking magnetic elements in magnetograms. Therefore, MFR formation by flux cancellation (see next section) remains a candidate.

We now pass into a discussion of observations of FC formation in the periphery of
ARs, rather in their cores, including quiet Sun (QS) filaments. AR periphery is indeed where most filament
channels form. Observations, \citep[e.g.,][]{gaizauskas1997,wang2007}, suggest that
such filament channels form during mainly interactions of emerging bipoles with existing
ones. The bipoles converge and flux cancellation and reconnection  take place.
These scant observations, along with the considerations of 
the solar-cycle characteristics
of filament channels within and outside ARs, with only
the latter showing solar-cycle dependence, led \citet[][]{mackay2010}
to conclude that emerging flux ropes are not relevant for the formation of filament
channels outside ARs, and "surface" phenomena, i.e., converge and cancellation
are more pertinent. 
We finally note here, that the observational inferences of bodily emerging
flux ropes within AR cores discussed in the previous paragraphs, may not be attainable for the
case of filament formation outside ARs, given the weaker (horizontal) magnetic fields in these areas.

Flux emergence in multipolar ARs can eventually lead
to the formation of MFRs \citep[][]{chintzoglou2019}. These authors reported observations of multiple bipoles, emerging either simultaneously
or sequentially, in two emerging ARs. The
collision between oppositely signed nonconjugated polarities
(i.e., polarities not belonging to the same bipole)
of the different emerging bipoles within the same AR gave rise to shear
and flux cancellation, hence this process was called  
collisional shearing. Photospheric flux cancellation and
reconnection above the photosphere in the PIL(s) undergoing
collisional shearing progressively converted SMAs into pre-eruptive MFRs and produced intense flare clusters for the duration of collision. 
In the same study, a data-driven evolutionary magneto-frictional
model \citep[e.g.,][]{fisher2015}
was applied to time-series of photospheric vector
magnetic field and Doppler measurements of the
two analyzed ARs, and it was able to capture and reproduce
the various stages of collisional shearing, including
the progressive conversion of SMAs to pre-eruptive MFRs.
In addition, using Doppler observations it was determined that the active part of the AR PIL showed downflows, further supporting that the pre-eruptive MFRs did not emerge bodily and were produced by strong cancellation.

\subsubsection{Outstanding Issues}
We conclude this section with a list
of pending issues with regards to flux emergence
simulations and observations:  \\
- improve the physical realism of flux emergence
simulations; \\
- lack of systematic studies of full and partial
flux-tube emergence ; \\
- lack of systematic calculations of twist in
flux emergence simulations; \\
- properties and eruptivity of bodily emerging
MFRs has not been studied yet; \\
- lack of extensive surveys looking for bodily emerging
MFRs ; \\
- lack of extensive surveys of FC formation at AR peripheries.

A detailed 
roadmap towards addressing these
issues is given in  Section \ref{s:path}.

\subsection{Flux Cancellation}
\label{ss:cancellation}

\subsubsection{Introduction}
Magnetic flux cancellation, whereby small-scale opposite magnetic polarities converge, collide and then subsequently disappear \citep[][]{martin1985} is a process ubiquitous all over the solar photosphere \citep[][]{livi1985}. Flux cancellation takes place along the PIL \citep[][]{babcock1955} that separates negative and positive photospheric polarities. In particular, cancellation can occur along the internal PIL of an AR, at an external PIL formed at the AR periphery, or in the QS.
According to the classical picture  of magnetic
flux cancellation described in
\cite{vanballegooijen89} it invariably leads to the formation of an MFR. 
A different model for magnetic cancellation is discussed in Section \, \ref{ss:condensation}, and when combined with helicity condensation, this type of cancellation leads to an SMA.

\subsubsection{Flux cancellation modeling} \label{sss:cance_mod}
Flux cancellation due to magnetic reconnection at the surface is typically modeled by modifying the lower boundary of a simulation domain that represents the photosphere. For example, in the work of \cite{Amari_1999,Linker_2001,Lionello_2001}, a potential-field extrapolation of either an idealized or observed AR magnetic field is used as the initial condition. This field is then energized by surface flows, such as shearing or twisting, followed by diffusion of the normal magnetic field on the surface, which is achieved by imposing an appropriate tangential electric field. The ensuing evolution produces a coronal MFR with dipped field lines that could support filament material. In the model of  \citet{devore2000p}, no specific surface flux cancellation is specified. Instead, the expansion 
due to footpoint motions which increase the pressure
of the sheared magnetic field in the corona  leads to magnetic reconnection, similar to the theoretical mechanism of \citet{vanballegooijen89}. Finally, in the models of \citet{Yeates_2008} and \cite{vanballegooijen00}, diffusion of the magnetic field at the coronal base (with the simultaneous absence of magnetic diffusion in the corona) concentrates magnetic field just above PILs, which becomes the axial field of the MFR formed by the subsequent reconnection.  

The complex physical mechanisms associated with flux emergence are reviewed in Section\,\ref{ss:emergence}. During the partial emergence of magnetic flux (where mass-laden portions of magnetic flux tubes remain rooted at the surface), the Lorentz force associated with expanding field lines that are able to drain material produces 
shear flows at and above the PIL \citep{manchester2004}. In addition, because of the rapid horizontal expansion of the magnetic field at the $\beta=1$ surface (see Figure \ref{fig2}), convergent flows can also be created by the self-consistent evolution of the magnetic field in the low atmosphere, leading to magnetic reconnection at those locations \citep{Mactaggart_2009}. Furthermore, the emergence of concave-up field lines (U-loops) can occur when the magnetic field becomes significantly distorted during its emergence in the turbulent convection zone \citep{Magara_2011}. Therefore, flux emergence can create the observational signatures of shearing and converging flows, and of flux cancellation. 

Recently, the ability to include into the simulations
the physical mechanisms of both flux emergence and 
flux cancellation due
to both magnetic-field evolution and magneto-convection has been developed.
By mimicking the radiative losses at the surface and including a thermal injection of energy at the base of the simulation domain, MHD simulations spanning a shallow convection zone to lower corona can now include reasonably realistic convective motions. \citet{Fang_2012} address the emergence of MFRs modified by the turbulent convection zone. Within the emerging structure, converging motions at the PIL, driven by a combination of magnetic-field evolution and granular motion, cause flux cancellation at the photosphere, which, along with tether-cutting reconnection in the corona, continues to build up sheared field lines in the corona. This type of study, performed on realistic solar AR timescales may help to understand what causes the flux cancellation observed prior to internal filament eruptions. 

Convective motions at the surface may also aid in the formation of filament channels both within and external to an AR, via the recent theoretical model of helicity condensation, discussed in Section\,\ref{ss:condensation}. 

\subsubsection{Flux cancellation observations}\label{sss:cance_obs}

The convergence of opposite magnetic polarities towards PILs that leads to flux cancellation is driven by the dispersion of decaying AR magnetic fields, through convection and emergence of AR bipoles into pre-existing magnetic field, and also by the collision of bipoles during the emergence of multipolar ARs. The process of flux cancellation is often observed leading up to the formation of filament channels, filaments and CMEs \citep[][]{martin1985,martin2012,martin1998,gaizauskas2001,gaizauskas2002,wang2007,mackay2008,mackay2014,chintzoglou2019}, suggesting that flux cancellation plays an important role in the construction of pre-eruptive magnetic-field configurations.
Small-scale brightenings and jets have been observed in the corona in connection with cancellation sites \citep[e.g.,][]{wang2013}. However, such brightenings are most commonly observed in H$\alpha$ and He II, with fainter or no signatures in coronal emissions.

There are currently three proposed scenarios that describe the physical processes that lead to flux cancellation \citep[][]{zwann1987}: U-loop emergence \citep[][]{vandriel2000,bernasconi2002}, $\mathrm{\Omega}$-loop submergence \citep[][]{harvey1999,chae2004,yang2009,takizawa2012}, and magnetic reconnection taking place low in the solar atmosphere followed by $\mathrm{\Omega}$-loop submergence \citep{vanballegooijen89}. Current observations cannot
easily distinguish between these possibilities. 

In the case that flux cancellation is associated with magnetic reconnection,  as in the  \cite{vanballegooijen89}
model, then sheared magnetic fields reconnect in the PIL leading to the formation of twisted field lines pertinent to a MFR. More specifically, the sheared field evolves from an SMA to two sets of loops that form a ``double-J'' shape with a low-emission channel between them. The low emission channel is expected to contain cooler plasma (about 1MK), but this has not yet been detected. From this configuration, the inner end points of the two J’s merge and a continuous S-shape forms, but this may also be a projection effect. This continuous S-shape corresponds to a SXR or extreme ultraviolet (EUV)  sigmoid.
\citet[][]{green2014} show this evolutionary scenario for four ARs, with the transition from SMA to double-J to continuous S-shape typically taking a couple of days.
The continuous S structure is highly supportive of the presence of an MFR with helical field lines
with around one turn, where heated plasma is  confined.
However, the MFR may also be present at the time of the two J's. The double-Js then represent remnant SMA field that is wrapping around the forming MFR but which has not yet undergone reconnection and been built into the MFR.  

During flux cancellation  an amount of flux equal to the amount removed from the photosphere is available to be built into an MFR.   Therefore, flux cancellation observations present a way to investigate how much magnetic flux has been built into an MFR before its eruption and we hereby discuss such estimates. Previous observational studies of flux cancellation monitoring the
magnetic flux evolution in the photosphere, suggest that around 30-50\% of the AR flux cancels on average 2-4 days prior to the occurrence of a CME \citep[][]{green11,baker2012,yardley2016,yardley2018a}. However, these values represent an upper limit as the amount of flux that is built into an MFR is dependent on the AR properties, such as the shear of the magnetic field and the length of the PIL along which flux cancellation is taking place \citep[][]{green11}.

In addition, static or time-dependent 
(non-linear force-free field;
NLFFF) models provide an alternative method to probe how much flux is contained in an MFR before its eruption. The models include the flux rope insertion method \citep[][]{vanballegooijen04} to produce a NLFFF extrapolation or the NLFFF evolutionary model of \citet[][]{mackay2011}, to model pre-eruptive configuration of the magnetic field. Using these models it is possible to investigate the ratio of axial and poloidal (i.e., at planes perpendicular to the axis ) 
flux in the MFR to that of the overlying field for stable and unstable MFRs. Previously, the limit for the axial flux in an MFR that can be held in stable force-free equilibrium by the overlying field of the AR was found to be up to 10-14\% \citep[][]{bobra2008,su2009,savcheva2009}. On the other hand, more recent studies that combine both observations and models have suggested that 20-50\% of the AR flux could be contained in a stable MFR \citep[][]{savcheva2012,gibb2014}. There may be also poloidal magnetic flux added to the MFR via reconnection in the transition region or corona, but lack of routine magnetic field observations at these layers prevents any direct assessment of it.

For sigmoids that form along the internal PIL of decaying ARs, the coronal field appears to follow a systematic evolution when viewed in SXR observations. These regions match closely the mechanism of \cite{vanballegooijen89} in that they exhibit flux cancellation along the PIL and an increasingly sheared field in the corona. A recent study by \cite[][]{Savcheva&al2014} has shown that the eruptive activity of 72 sigmoidal ARs is more strongly correlated with flux cancellation than with emergence. The study found that 57\% of the sigmoids were associated with flux cancellation compared with only 35\% occurring as a result of flux emergence. The rest of the regions showed a fairly constant magnetic flux evolution.

\begin{figure*}[h]
\centering
\includegraphics[width=0.8\textwidth]{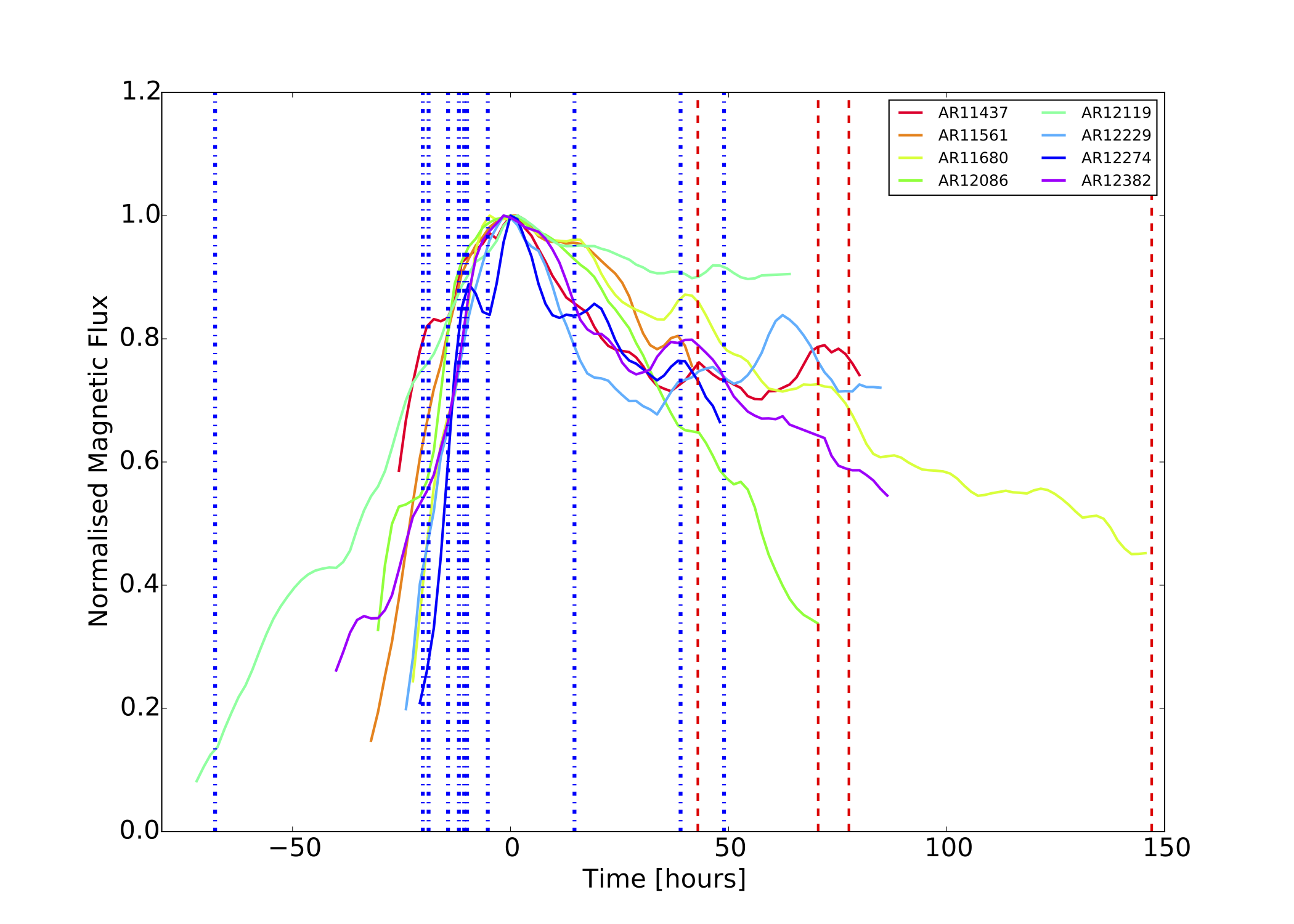}
\caption{
The temporal evolution of normalized magnetic flux in CME-producing ARs studied by \citet[][]{yardley2018a}. The vertical lines indicate the timing of CME eruptions from PILs that are internal to the AR (red dashed lines) and PILs that are external to the AR (blue dot dashed lines). CMEs from external PILs tend to happen during the flux emergence phase (t $<$ 0 hr) while CMEs from ARs' internal PILs happen well after the emergence phase (t $>$ 0 hr).
\copyright AAS. Reproduced with permission.}
\label{fig3}
\end{figure*}

 A secondary effect of flux cancellation is that it reduces the magnetic flux that contributes to the overlying, stabilizing field of the MFR. If enough flux is cancelled from the overlying field and incorporated into an MFR, a force imbalance occurs, which leads to a catastrophic loss of equilibrium and a CME \citep[e.g.,][]{lin2000,bobra2008}. Or, if the overlying field of the AR decays rapidly enough with height, the flux rope can become torus-unstable \citep{kliem06,torok07,demoulin10,kliem14}. Previous observational studies of flux cancellation have found a ratio of flux contained in the MFR compared to the overlying arcade of 1:1.5 \citep[][]{green11} and 1:0.9 \citep[][]{yardley2016}. \citet[][]{yardley2018a} recently conducted a comprehensive study of the CME productivity in a sample of 20 bipolar ARs in order to probe the role of flux cancellation as a CME trigger. The magnetic flux evolution was analyzed during the full AR evolution spanning from emergence to decay. This study found that the ratio of flux cancelled available to be built into an MFR before eruption compared to the remaining, overlying field was found in the range 1:0.03 to 1:1.57
for ARs that produced low-altitude CMEs originating from the internal PIL. The small ratios imply that the assumption that the amount of flux cancelled is equal to the amount of flux injected into the flux rope does not necessarily apply here. This needs further investigation. They suggest that a combination of the convergence of the polarities, magnetic shear and flux cancellation are required to build a pre-eruptive configuration and that a successful eruption depends upon the removal of a sufficient amount of the overlying field that stabilizes the configuration. The study also showed that the type of CME produced depends upon the evolutionary stage of the AR. CMEs originating above external PILs (between the periphery of the AR and the QS) occurred during the emergence phase of the AR, whereas, CMEs originating above an internal PIL of an AR occurred during the region's decay phase (see Figure \ref{fig3}).

Note that the majority of the observational studies of flux cancellation discussed above refer to bipolar and decaying
ARs. Flux cancellation can also occur in multipolar and emerging ARs (\citet[][]{chintzoglou2019} 
and discussion of  Section \ref{sss:fe_obs}). We note
that measuring flux cancellation in periods of 
flux emergence is challenging (in comparison to measuring cancellation in decaying ARs) because the magnetic flux is increasing during the emergence phase.
By tracking 
opposite polarity footpoints and the flux balance of multiple emerging bipoles
within the same AR undergoing collisional shearing (see Section 
\ref{sss:fe_obs}) it was found by \citet{chintzoglou2019}
that amounts of magnetic flux of up to $\approx$ 40 $\%$ of the net magnetic flux of the smaller emerging bipole may be cancelled. The reported collisional shearing and magnetic cancellation preceded major eruptive activity in the analyzed ARs.

\subsubsection{Outstanding Issues}
We conclude this section with a list
of pending issues with regards to flux cancellation
modeling and observations:\\
- there are multiple physical processes that could be associated with flux cancellation and it is
not clear which dominates (if any); \\
- the exact amount of flux that builds into the pre-eruptive magnetic configuration 
associated with flux cancellation is not fully constrained; \\
- the exact atmospheric height at which the magnetic reconnection that could be associated with flux cancellation occurs is still unknown; \\
-it is difficult to follow the full magnetic field evolution of ARs from emergence to decay, particularly for large or complex ARs, due to current observational limitations; \\
-flux cancellation during episodes of flux emergence needs to be taken into account.

A detailed roadmap towards addressing these issues is given in  Section \ref{s:path}.
%

\subsection{Helicity Condensation}
\label{ss:condensation}

\subsubsection{Introduction}
In recent years, a third mechanism has been proposed for the formation of filament channels, the so-called condensation of magnetic helicity at PILs \citep[][]{antiochos2013}. Magnetic helicity is the topological measure of twist, writhe, and the linkage of magnetic field lines \citep[e.g., the review of][]{pevtsov2014}. An attractive property of magnetic helicity is that it is conserved even under  magnetic reconnection \citep[e.g.,][]{berger1984}. 
Furthermore, for a turbulent system, helicity has been shown to undergo an inverse cascading toward larger spatial scales.
The helicity condensation model implies that
magnetic shear accumulates due to this helicity cascading in a
given coronal flux system. 
The primary assumption  is that a net helicity is injected into every coronal flux region 
by the  continuous large and small-scale photospheric motions and 
associated flux emergence and submergence through the photosphere. 
It has been long observed that large-scale magnetic structures in the corona, such as sunspot whirls, exhibit a pronounced hemispheric helicity preference 
that was first postulated by \citet{Seehafer90}
and assesed in photospheric magnetic field observations by \citet{Pevtsov95} (see \citealt[][]{pevtsov2014} for more recent references). This preference is especially strong ($>$90\%)  for erupting filaments \citep[][]{ouyang2017}. 
A secondary requirement for the helicity-condensation model is that the coronal magnetic field is constantly undergoing turbulent-like reconnection, as in the nanoflare model \citep[][]{parker1983,klimchuk2006} for coronal heating. Given the observed constant small-scale convective flows and flux emergence/cancellation at the photosphere, it seems inevitable that the corona must be in a state of turbulent reconnection, irrespective of whether this process is important for coronal heating.  Such resistive MHD turbulence is expected to produce an inverse cascade in the helicity due to helicity conservation even under non-ideal processes.

\begin{figure*}[h]
\centering
\includegraphics[width=\textwidth]{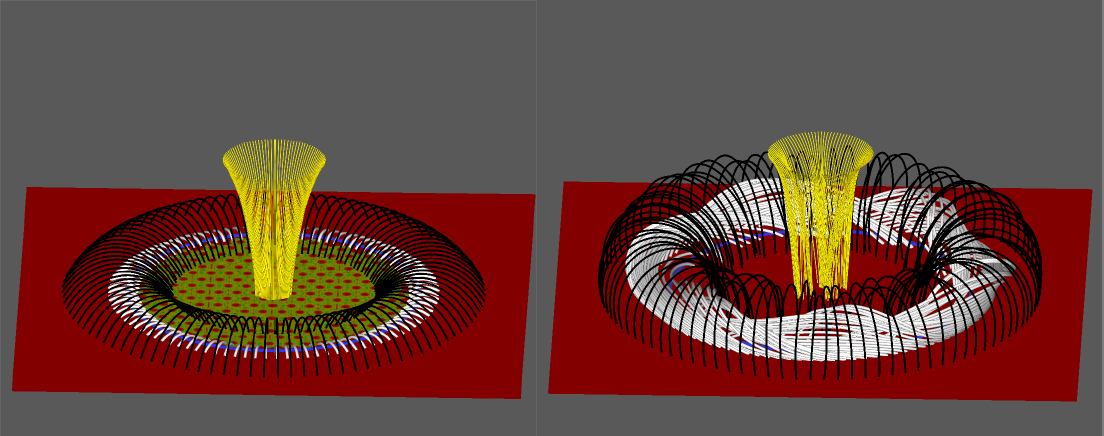}
\caption{Initial (left) and final (right) states of a coronal system driven by random small-scale helicity and free energy injection. The blue circle indicates the PIL and the contours inside this PIL in the left panel show the flow pattern at the photosphere. The pattern is rotated randomly after a half rotation \citep[From][]{knizhnik2017}. \copyright AAS. Reproduced with permission.}
\label{fig4}
\end{figure*}

Inverse cascading is a well-known phenomenon in laboratory plasmas, and generally leads to the “self-organization” of a turbulent system \citep[][]{biskamp1993}. The helicity condensation model argues that for the corona self-organization manifests itself as the formation of sheared filament channels about PILs. Simulations by \citet[][]{knizhnik2017} demonstrate the basic mechanism of the model. The left panel of Figure \ref{fig4} shows the initial coronal system consisting of a potential magnetic field due to a bipolar flux distribution at the bottom plane (the photosphere), with a circular positive polarity region surrounded by a negative polarity. The blue circle in Figure \ref{fig4} indicates the PIL. Three sets of field lines are shown: yellow lines that are open so that any stress by the photosphere simply propagates outward, black corresponding to typical coronal loops, and low-lying white lines very near the PIL. This initially zero helicity system was driven by small-scale free energy and helicity injection due to a simple photospheric flow pattern consisting of multiple rotations. The rotations filled the positive polarity region, thereby driving all the flux in the system. To reproduce the stochastic nature of photospheric convective flows, the rotations were randomly shifted about the bottom plane after every half-turn or so.

The right Panel of Figure \ref{fig4} shows the end state of the system after approximately 90 rotations and demonstrates a striking example of self-organization. As expected, the yellow open field lines are relatively unchanged, but the black lines in the central portions of the flux system also show negligible shear or twist even though most of the helicity and free energy has been injected into this region. The white field lines near the PIL, on the other hand, are all highly sheared along the PIL. Note that the shear is highly localized about the PIL,
as in observations \citep[e.g.][]{schmieder1996} and exactly
as required for a filament channel. Furthermore the black
  lines, although pushed upward by the increased magnetic pressure of
  the filament channel,  are smooth and laminar,  as
observed for coronal loops \citep[][]{schrijver1999}.

Previous studies discuss the buildup of shear flows due to a significant Lorentz tension  force. This is due to the non-neutralized currents developed exclusively along strong PILs \citep[][]{Georgoulis12}, verified independently by \citet{torok14a}, although the latter study does not necessarily attribute net currents to the Lorentz force. The entire process leading to Lorentz force-driven shear is  ultimately fueled by flux emergence \citep[][]{manchester2004}. Shear flows largely cease when flux emergence is complete, even at established PILs.

A key point evident from Figure \ref{fig4} is that the FC field lines
show no evidence for a coherent global twist. The structure is clearly
that of a SMA rather than a MFR. For a SMA, the helicity in the system
is primarily due to the mutual linkages between the strongly sheared
FC field and the overlying quasi-potential flux. Figure \ref{fig4} is
a dramatic demonstration of how reconnection can convert one form of
geometrical linkage to another. The helicity injected into the system
was all due to small rotations, in other words, the internal twist of
many small flux tubes. Reconnection, however, converts this helicity
into a global shear. But we caution the reader that inverse
cascading of helicity does not necessarily lead to shear as shown
amply, for example, by studies of the kink instability. In addition,
a strongly stressed PIL is required to obtain substantial shear 
\citep[see also the theoretical work by][]{alexakis2006}. We also note
that the final system in Figure \ref{fig4}, is not in a minimum energy 
linear force-free state. The reason is that the line-tying introduces 
additional constraints on the system evolution beyond only global helicity 
conservation \citep[][]{antiochos2013}.

\begin{figure*}[h]
\centering
\includegraphics[width=\textwidth]{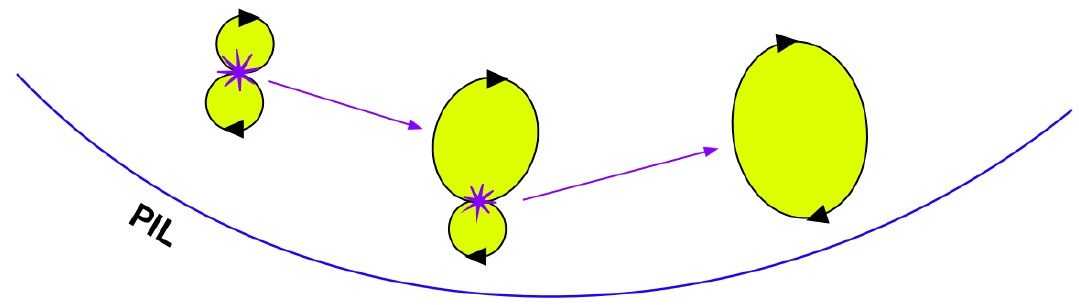}
\caption{Schematic drawing of the evolution giving rise to the results
  of Figure \ref{fig4}. The yellow circles indicate the photospheric
  footpoints of twisted coronal flux tubes. Reconnection of the twist
  component of the two elemental flux tubes on the left, which were
  formed by photospheric motions, causes them to merge and form the
  larger twisted flux tube in the center. Reconnection of this flux
  tube with another elemental flux tube produces the large flux tube
  on the right. Note that this process successively moves magnetic
  stress toward the PIL where it appears as a large-scale coherent shear.} 
\label{fig5}
\end{figure*}

A simple schematic as to why coronal helicity condenses onto the PIL is presented in the Figure \ref{fig5}. Since the photospheric driving speeds, $\sim$\,1 km/s, are much smaller than the coronal Alfv\'en speed, and the corona is low $\beta$, $\sim\,{10}^{-2}$, the coronal evolution is approximately given by a quasi-static sequence of force-free states. In that case the twist field 
i.e., the azimuthal field component
is constant along a twisted flux tube, neglecting the effect of flux-tube expansion, and consequently, the evolution can be captured by considering only the twist field projected onto the 2D photospheric surface. Figure \ref{fig5} shows that reconnection between any two neighboring twisted flux tubes results in the spreading of the twist field so that it encompasses the flux of both tubes. Further reconnection of these larger tubes spreads the twist flux further until it “condenses” against the PIL, because this boundary defines the extent of the flux system. 
This shows that inverse cascading within the PIL flux system is at the heart of the helicity condensation model with reconnection being a key process here. Rigorous numerical calculations, such as the one shown in Figure \ref{fig4} above \citep[e.g.,][]{knizhink2015,knizhnik2017,zhao2015}, have confirmed both the basic picture of Figure \ref{fig5} and the estimates for the rate of FC formation given in \citet[][]{antiochos2013}.

\subsubsection{Theoretical Issues}
At least two important theoretical issues remain to be addressed by
the helicity-condensation model. The first is the effect of a
multi-scale photospheric driving. The calculations to date used a
single scale for both the helicity injection and the small-scale
driving that powers the turbulent reconnection, but in the corona
these are likely to occur at different scales. 

The second critical theoretical issue is the effect of flux
cancellation on helicity condensation. In the \citet[][]{knizhnik2017}
simulation, the PIL itself was not driven in order to avoid the
effects of flux cancellation, but there is no doubt that flux does
cancel at PILs. As originally described by 
\citet[][]{vanballegooijen89}, 
and subsequently modeled by many authors 
\citep[e.g.,][]{amari2000,mackay2010}  
flux cancellation
  occurred systematically along a PIL in a quasi-2D evolution. In this
  classic scenario, flux cancellation at the photosphere is physically
  similar to flare reconnection along a quasi-2D X-line. Both
  simulations and observations have shown that such quasi-2D
  reconnection inevitably produces a near potential arcade below the
  reconnection line (the flare loops) and a large-scale MFR with
  global systematic twist above the line (the CME ``plasmoid''). 

\begin{figure*}[h]
\centering
\includegraphics[scale=0.3]{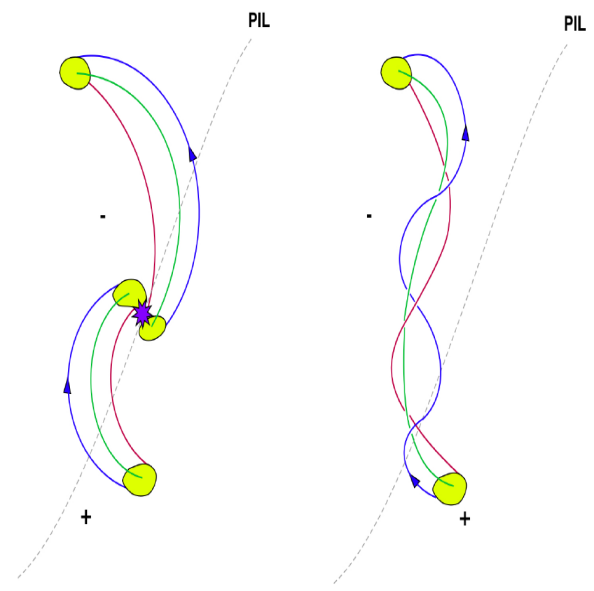}
\caption{Schematic drawing of realistic 3D flux cancellation at a large-scale photospheric PIL. Shown on the left are four photospheric flux concentrations (yellow patches) that are the footpoints of two coronal flux tubes. Originally all four concentrations were far from the PIL, but photospheric motions bring two opposite polarity concentrations to the PIL where they cancel by reconnection and submergence. This cancellation results in the single coronal flux tube shown on the right, which is twisted and has both footpoints far from the PIL.}
\label{fig6}
\end{figure*}

Observations show, however, that except possibly at the
centers of young ARs where flux has just emerged, flux cancellation is
far from systematic, and is instead a quasi-random fully 3D process.
High-resolution observations of the photospheric magnetic field show
that cancellation consists of two strong concentrations of opposite
polarity flux (``magnetic elements''), seemingly randomly colliding
and disappearing \citep[][]{berger1996}.  Figure \ref{fig6}
illustrates the nature of flux cancellation given the observed
photospheric flux evolution. Two concentrations of opposite polarity
flux are very near a large-scale PIL, where they collide and cancel by
photospheric motions. The other ends of the two flux tubes are not
equally near to the PIL, as in the systematic cancellation models, but
instead are much further away. Even this picture is a gross
simplification. In the actual corona, the concentrations near the PIL
would connect to several concentrations away from the PIL, and there
is always weak flux between the concentrations
\citep[e.g.,][]{wiegelmann2014,cheng2016}.

The implication of Figure \ref{fig6} above is that flux cancellation is simply another source of small-scale twist injection into coronal loops, just like convective flows and flux emergence.  This
small-scale twist, along with the twist from the other sources, then
undergoes the helicity cascade process and ends up as a large-scale
PIL shear. This small-scale twist, along with the twist from the other sources, then undergoes the helicity cascade process and ends up as a large-scale PIL shear. Rigorous MHD simulations
are now required to test this hypothesis in tandem with observational
studies aiming to fully characterize the properties of flux
cancellation on both small and large spatial scales.

\subsubsection{Observations of magnetic helicity}

\begin{figure*}
\centering
\includegraphics[width=\textwidth]{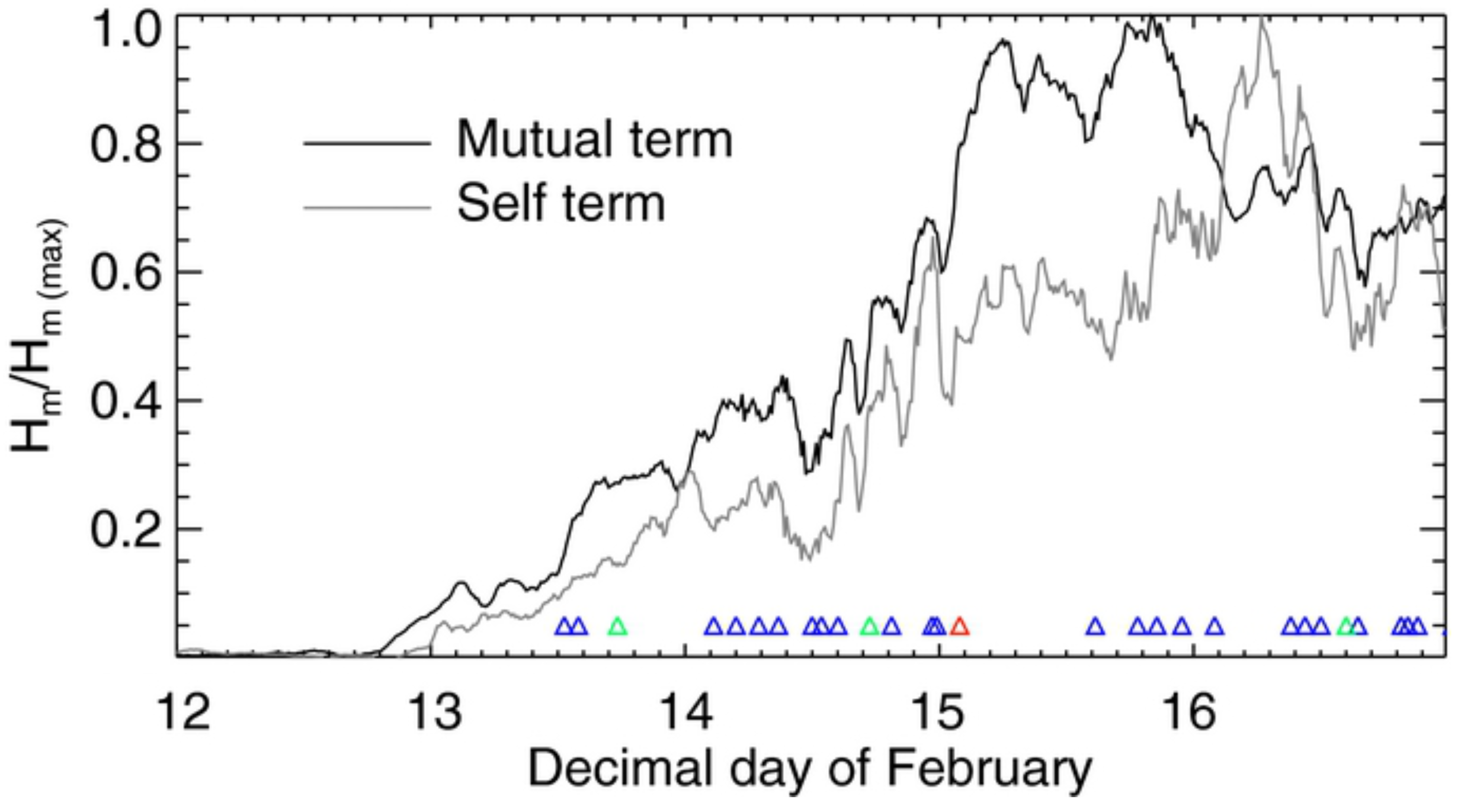}
\caption{An example of mutual-to-self relative magnetic helicity
transformation during the emerging-flux phase in observed NOAA AR
11158, primarily along the AR’s strong PIL. The self-helicity term
(gray curve) follows closely the mutual-helicity one (black curve),
lagging from it by a few to $\sim$24 hours. The triangles adjacent
to the time axis correspond to onset times of flaring activity in
the AR. Modified from Tziotziou et al., (2013).
\copyright AAS. Reproduced with permission.}
\label{fig:m2s}
\end{figure*}

Helicity exhibits inverse cascade to shorter wavenumbers, a fact well-established from theory and simulations \citep[][]{alexakis2006}. However, most pieces of evidence about its consequences in the solar atmosphere are qualitative. Apart from the concentration of shear almost exclusively around flux-massive PIL areas \citep[e.g.,][]{hagyard1984,moore1987,zirin1987,tanaka1991,
schrijver2007},  indirect evidence for
the inverse cascade of helicity is provided by the conversion of some of the twist of MFRs to writhe 
\citep[a measure of the deformation of the MFR axis; see, e.g.,][]{torok10} in the course of the helical kink instability, which has a threshold that is reached at around a twist of 
$N=1.25$
or more \citep[e.g.,][]{Hood&Priest1981,torok04}. We note that in the line-tied corona the kink by itself could not move helicity across field lines, but it would enhance the possibility of reconnection, which then transports helicity. Furthermore the amount of writhe produced depends primarily on the 
evolution
of the instability, 
not just on
the initial twist \citep{torok14b}.
A typical observed example of a helical kink instability can be found  in \citet[][]{rust2005}. Moreover, the inverse cascade of helicity can be invoked to interpret the gradual mutual-to-self helicity conversion pattern reported by \citet[][]{tziotziou2013} (Figure \ref{fig:m2s}) in an emerging eruptive AR  
\citep[see also][for another example]{Li2014}. 
Inverse helicity cascading has been quantified in phase space by
\citet[][]{zhang2014,zhang2016}, who found that the unsigned current helicity spectrum, $Hc(k,t)$, of two ARs show a ${k}^{-5/3}$ power law. They also found that the current helicity spectrum is related to the magnetic helicity spectrum, $Hm(k,t)$, via $Hc(k,t) \simeq {k}^{2} Hm(k,t)$). This result is consistent with simulations of hydromagnetic turbulence \citep[e.g.,][]{brande2005} and implies that turbulence becomes gradually less helical towards smaller scales. The same authors studied the solar cycle evolution of helicity spectra and found that helicity spectra steepen as we progress to solar maximum, emphasizing the large-scale magnetic field. 

Indirect observational evidence in support of the helicity-condensation model comes from the studies of filament chirality  \citep[e.g.,][]{mackay2005,yeates2012}. These authors found that the standard processes of differential rotation, AR emergence, flux diffusion and cancellation, etc. could not reproduce the observed global distribution of filament chirality and eruption rate, especially during the declining phase of a solar cycle.  The implication from these studies is that an additional mechanism is needed for injecting helicity into the corona and concentrating it along PILs. The helicity condensation model provides such a mechanism \citep[][]{mackay2018}. 

Several observations support the important role of helicity in the initiation of eruptions (see \citealt{pevtsov2014} for details). \citet[][]{nindos2004} found that, statistically, the pre-flare absolute value of the linear force-free field parameter and the resulting coronal helicity are larger in ARs that produce eruptive major flares than in those that produce major but confined flares. A similar conclusion was reached by \citet[][]{labonte2007} who calculated the helicity flux in large samples of X-flaring and non-X-flaring ARs and found that a necessary condition for the occurrence of X-class flares is that the peak helicity flux has a magnitude $> 6\times{10}^{36} \mathrm{{Mx}^{2}{s^{-1}}}$. \citet[][]{park2008,park2010}  studied about a dozen X-class flares and found that they were preceded by a significant helicity accumulation, (1.8-32)$ \times {10}^{42} \mathrm{{Mx}^{2}}$, over periods of half a day to a few days.

Using the method developed by 
\citet[][]{georgoulis2012hm}, \citet[][]{tziotziou2012} calculated the instantaneous relative helicity and free magnetic energy of several ARs and found a significant monotonic correlation between the free energy and the relative helicity. This correlation reinforces the notion that, in addition to free energy, helicity may play a central role in solar eruptions. They also found that the eruptive ARs were well segregated from the non-eruptive ones, in both free energy and helicity, with free energy and helicity thresholds for the occurrence of major (GOES M-class and above) flares of $4\times {10}^{31}$ erg and 2 $\times {10}^{42} \mathrm{{Mx}^{2}}$, respectively. \citet[][]{nindos2012} found that the initiation of eruptions in a large emerging AR depended critically on the accumulation of both free energy and helicity in the corona and not on the temporal evolution of the variation of the background field with height.

\citet[][]{pariat2017} performed eruptive and non-eruptive 3D MHD simulations in which they decomposed the relative helicity, $H$, into a term $H_j$ associated with the current-carrying magnetic field  and a term $H_{pj}$ associated with the intra-helicity between potential and current-carrying fields. They found that the ratio $H_j/H$ yields a reliable eruptivity proxy. It will be interesting to check the performance of $H_j/H$ against observations, as well, although one will ultimately need to rely on a model or extrapolation for the unknown coronal magnetic field. The above results highlight the importance of helicity buildup in the initiation of eruptions. Along with other observational and modeling results, reviewed by \citet{georgoulis2019}, such findings place magnetic helicity on equal footing with free magnetic energy in terms of their roles in the initiation of solar eruptive events.

\section{The Magnetic Configuration of Filament Channels}
\label{s:configurations}

We now discuss the theoretical expectations and observational signatures used in interpreting the FC magnetic configurations as SMAs or MFRs. Although we describe the two types as separate subsections, we contrast the SMA and MFR interpretations throughout. To make the discussion easier to follow, we organize the text according to the various signatures (e.g. prominence dips, sigmoids, cavities, etc.), given in \textit{italics\/} at the beginning of the relevant paragraph.

\subsection{Sheared Magnetic Arcades (SMAs)}
\label{ss:SMA}
\subsubsection{Introduction}
 Early attempts of modeling the magnetic structure of filaments or prominences were based on 2D or 
2.5D (translationally invariant) arcade configurations. 
None could account for the existence of plasma-supporting magnetic dips.
On the one hand, bipolar 2.5D force-free arcades were unable to generate magnetic dips \citep[see][]{klimchuk1990,amari1991}, which were considered necessary to support filament material. On the other hand, MHD models had difficulty in forming low-$\beta$ and stable weight-induced dips \citep[see e.g.][]{wu1990,choe1992}. 
So SMAs with dips, if they ever existed, had to be intrinsically 3D, unlike MFRs that already have dips in 2.5D. This was not only a geometrical challenge, but also a computational one, given the limitations in the early 90's. A second challenge was on the observational side. Magnetic field measurements within prominences were showing a vast majority of so-called inverse-polarity (IP) configurations  \citep[][]{leroy1984,bommier1998}.  In these configurations, the horizontal field viewed from above points toward the opposite direction than the regular normal-polarity (NP), i.e., it goes from the negative toward the positive polarity. While this peculiar behavior is readily satisfied in the windings of MFRs, it was not obvious how this could be achieved in mere bipolar arcades, which are naturally NP 
\citep[][]{klimchuk1988}. 

\begin{figure*}[h]
\centering
\includegraphics[scale=0.2]{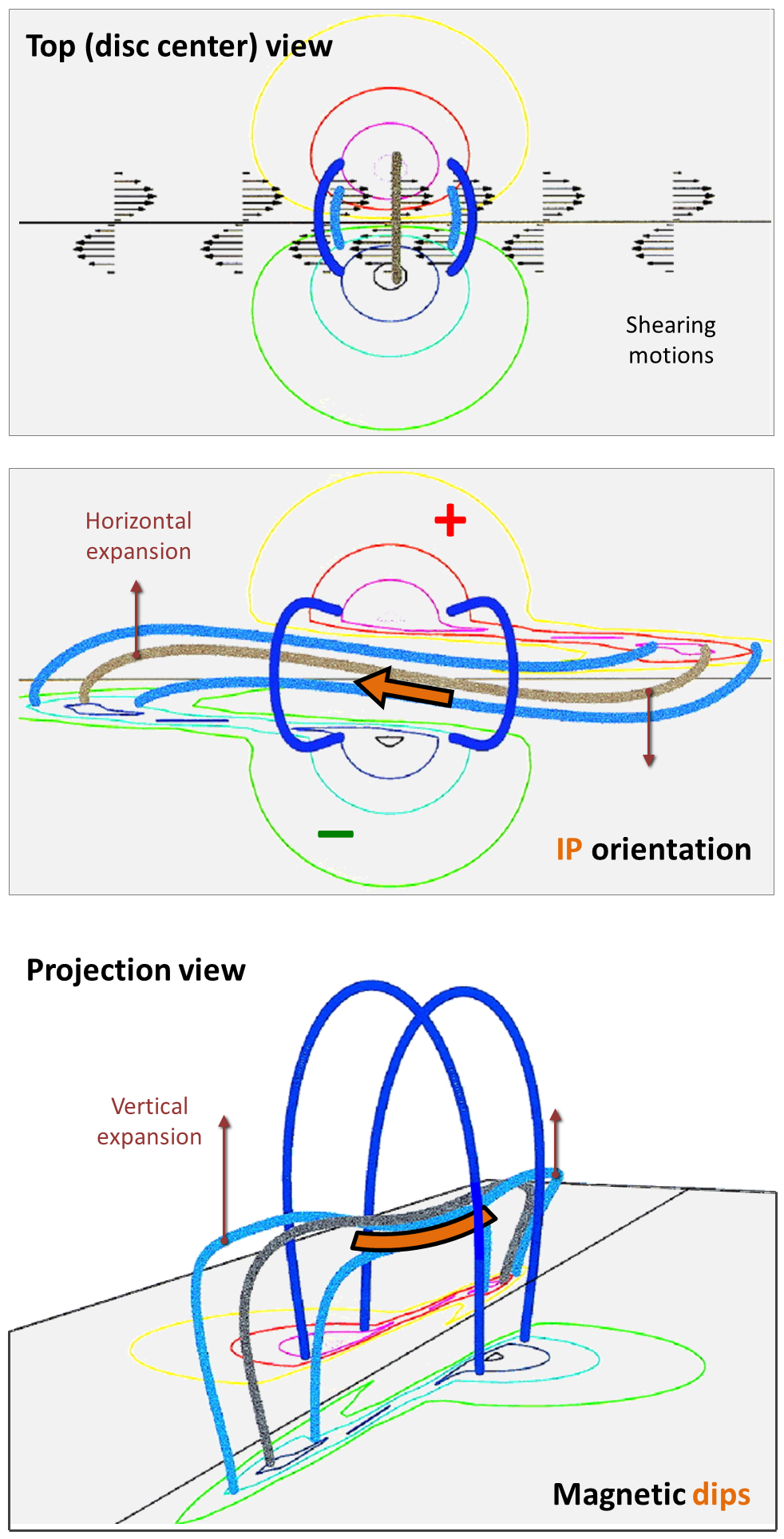}
\caption{Overhead views of the initial condition  (upper panel) and of the  final state  (middle  panel) of a SMA simulation.  Perspective view of the final state of the same simulation (lower panel). The colored contours on the bottom of each panel correspond to photospheric Bz-contours while the thick colored lines correspond to magnetic field lines. Modified from \citet[][]{devore2000p}.\copyright AAS. Reproduced with permission.}
\label{fig7}
\end{figure*}

All these challenges were simultaneously solved in \citet[][]{antiochos1994} in terms of a fully 3D  model leading to formation of force-free and inverse-polarity dips in the SMA models. The computational limitations were overcome by using the original magneto-frictional approach from \citet[][]{yang1986} instead of a full-MHD model. The results were later confirmed by MHD modeling  (\citealt{devore2000p,aulanier2002}; Figure~\ref{fig7}). The geometrical issues were simplified to the extreme by concentrating surface-motion-induced magnetic shear close to and around the PIL, also satisfying observational constraints of strong shear being confined to PILs \citep[][]{schmieder1996}. It was found that when the extension across the PIL of the bipolar field is smaller than the length of sheared field lines along the PIL, then IP-dips are naturally produced (see Figure \ref{fig8}).

The first key result of this investigation was that the stretching of the sheared field lines away from these strong-field regions leads to their horizontal expansion towards weaker fields, i.e., towards the opposite side of the PIL. This produces S-shaped (sigmoidal) field lines with an IP orientation at their center. The second key result was that the vertical expansion of these sheared arcades is larger at large distances along the PIL than it is within the core of the strong bipolar fields, owing to stronger magnetic tension in the bipole center. This produces a dipped geometry at the center of the field lines, right where an IP orientation is present. 
 
The SMA model was also extended to interacting bipoles aligned along the PIL \citep[][]{devore2005,aulanier2006}. When the appropriate conditions are met, as far as observational constraints on magnetic helicity and prominence merging are concerned \citep[see][]{aulanier2008}, then the sheared bipoles reconnect with each other. We note that this reconnection takes place in the so-called finite-$B$ slip-running reconnection regime, i.e., occurring at regions of drastic changes in the magnetic field gradient, and does not abruptly change field-line connectivity \citep[][]{aulanier2006s}. Reconnection between sheared bipoles merely produces a longer SMA, with an inhomogeneous spatial distribution of dips that can support prominence plasma, as observed outside of ARs where the surface magnetic field is not concentrated in a single bipole.  Therefore, the SMA model can in principle account also for large quiescent prominences, either between or within decaying ARs, or in the polar crowns. 

The SMA model may be criticized because of its somewhat artificial distribution of flux at the Sun's surface, which results from the shear-driven elongation of the bipole along the PIL. While this issue remains to be addressed in single-bipole AR SMA models, solutions have been proposed in the framework of multi-bipole  quiescent-prominence SMA models (see below) and of the helicity-condensation model (see Section\,\ref{ss:condensation}).  Nevertheless, the SMA models
naturally provide a viable and straightforward model for the magnetic structure of FCs, i.e., for pre-eruptive magnetic configurations.

\subsubsection{Pre-eruption signatures of SMAs} \label{sss:SMA_pre}

\begin{figure*}[h]
\centering
\includegraphics[scale=0.25]{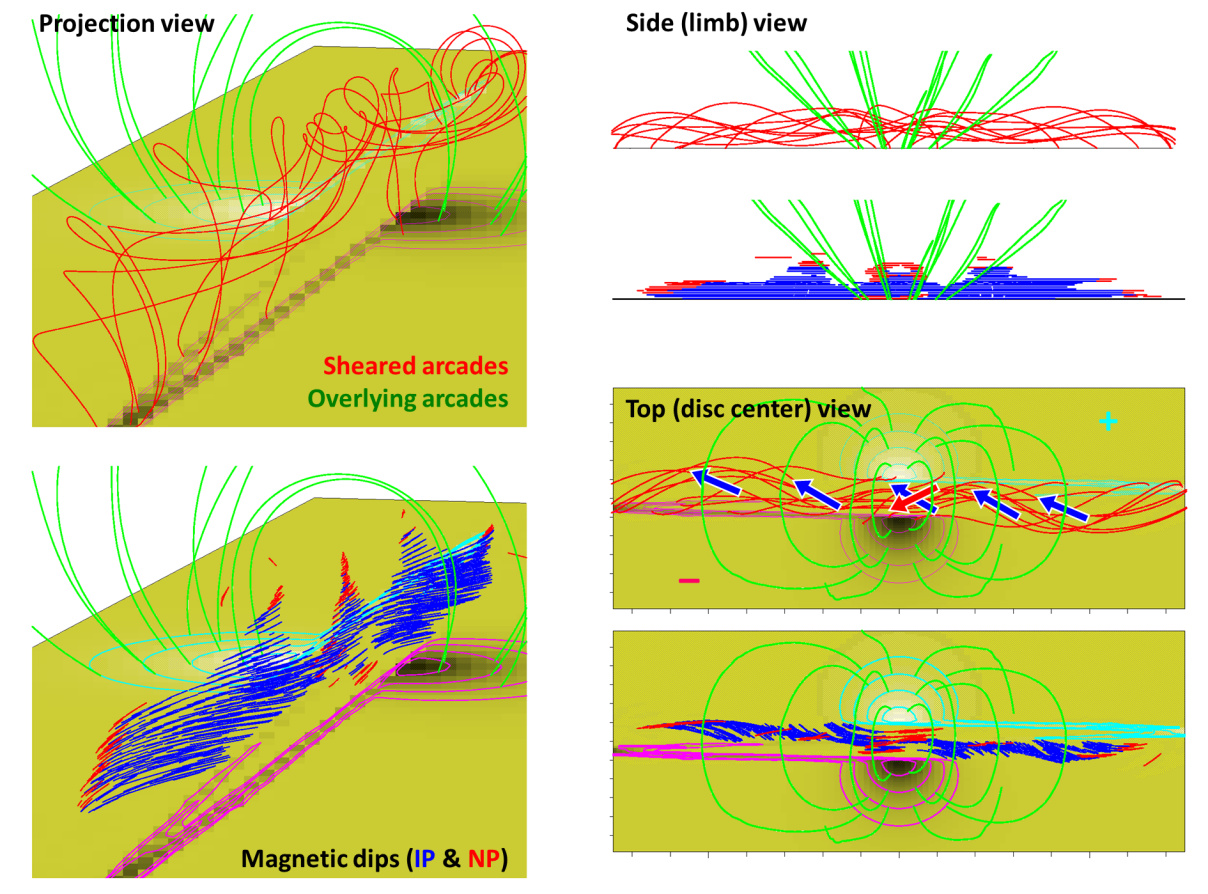}
\caption{Projection view of a SMA from a fully 3D MHD simulation applying large shears in the photosphere; SMA (red) and overlying arcade (green) field lines (upper left panel); distribution of IP (blue) and NP (red) dips (lower left panel). Side (limb) and top (disk center) views with the same same format as the projections views. Modified from \citet{aulanier2002}. \copyright AAS. Reproduced with permission.}
\label{fig8}
\end{figure*}

\textit{Magnetic Field Dips:\/} In the related MHD calculations, SMAs dominate the core of the PIL. Also, the modeled field-line dips are always dominated by IP-dips, irrespective of the shear magnitude. In addition, a key property is that all SMA models
discussed here possess narrow regions of NP dips at their top, within, and at the edge of the strong surface fields  (bottom left panel of Figure \ref{fig8}). When it was  identified, this property was proposed to be a clear-cut discriminator between SMA and MFR. This was disputed, however, when it was found that MFRs can also produce NP dips at their center, when they were perturbed by small-scale surface flux concentrations in the vicinity of a fragmented PIL \citep{aulanier2003}. 
Such small polarities hardly exist in the core of young solar ARs with $\delta$-spots
e.g., Fig. 17 in \citet[][]{vandriel2015}, from which the most energetic eruptions originate 
\citep[e.g., the review of ][]{toriumiawang}
. Therefore, \textit{identifying NP dips in compact AR sigmoids and filaments may provide a method for discriminating between the SMA and the MFR model.\/} 

SMA models tend to produce rather shallow dips. This comes naturally from differential shearing, which makes the SMAs expand differentially  in the horizontal and the vertical direction in the corona. Calculations of the formation and evolution of prominence condensations were achieved in such shallow dips, 
within the thermal non-equilibrium model \citep[][]{antiochos1991}. The calculations were done for generic shallow dips as well as nearly-flat field lines \citep[][]{antiochos1999,antiochos2000p,karpen2001,karpen2006}, and also using field line geometries that directly result from MHD simulations of single and double-dipole SMA models \citep[][]{karpen2006,luna2012}. 
A key property of SMA shallow dips (as well as their surrounding nearly-flat field lines) is that they permanently host very dynamic prominence condensations. This behavior is observed in many filaments \citep[][]{schmieder1991,schmieder2017,lin2003,lin2012}. This does not hold for highly twisted MFRs, where deep dips are produced by the large-scale winding of coronal field lines. 
In the SMA models, only the lower dips are deep enough to prevent a dynamic behavior. It has therefore been argued that this property is another possible discriminator between the SMA and the MFR models. In addition, shallow dips are also implied by the observation of linear threads in high spatial resolution images of prominences and filaments. \citep[e.g.,][]{okamoto2007,vourlidas2010}. 

Note that the dips lose their discriminating characteristic for \textit{weakly to
moderately twisted\/} MFRs i.e., up to two turns or so. 
In that case, the dips in the upper part of the MFR are shallow too
\citep[see, e.g., Figure 2e in][]{gibsonfan06}, leading to dynamic plasma condensations there \citep{xia2014,xia2016}, and stable condensations in the lower part of an MFR, just as for an SMA. The above discussion suggest that the dynamic behavior of prominence plasmas rules out \textit{highly twisted MFRs\/} as viable models for quiescent prominences and hence, more generally, as CME progenitors. 

One has to take into account, though, that prominence models based on the distribution of field-line dips \citep[e.g.,][]{aulanier99,aulanier2002,lionello2002,vanballegooijen04,dudik2008,zuccarello16}, as well as MHD simulations of prominence formation in MFR configurations \citep{xia2014,kaneko18,fan18} suggest that prominence plasma is often  organized in a vertical, sheet-like structure, as the observations suggest for quiescent prominences. In such cases, prominence threads would occupy only relatively short sections of twisted field lines \citep[since such field lines cross the sheet rather than running along it; see, e.g., Fig.\,3 in][]{xia2016}, rendering the display of twist in quiescent prominences improbable.

\begin{figure*}[h]
\centering
\includegraphics[scale=0.25]{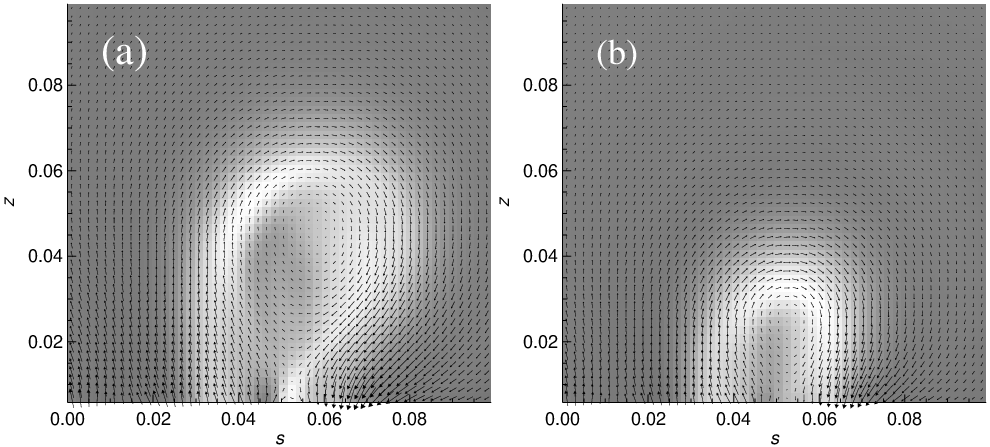}
\caption{Force-free parameter $\alpha$ (gray scale)
and in-plane magnetic field vectors (arrows)
in a vertical plane perpendicular to the magnetic axis of two numerically relaxed NLFFF models of an AR-FC, constructed using the flux rope insertion method. In (a), a higher amount of axial (i.e., highly sheared) flux was inserted, resulting in a so-called hollow-core MFR  (Section~\ref{sss:MFR_theory}), which features an HFT at its underside and is slightly unstable. In (b), the lower amount of axial flux results in a stable hybrid configuration, close to a hollow-core MFR, which neither contains 
an HFT nor a BP. Modified from \citet{Kliem&al2013}. \copyright AAS. Reproduced with permission.}
\label{fig91}
\end{figure*}

Moreover, some FCs may consist of so-called 'hollow-core' MFRs, in which twisted field lines exist only at the periphery of the configuration, while the core field lines run more or less parallel \citep[e.g.,][see also Fig.\ 
\ref{fig91}]{bobra2008,titov14}. 
Thus, the fact prominence threads very rarely show indications of twist does not exclude that the underlying structure is an MFR.  

\textit{Magnetic Field Topology:\/} Topology is a fundamental difference between SMAs and MFRs.  MFRs are separated from the surrounding coronal arcades. In 3D, this requires either a coronal separatrix surface, which emanates from a bald patch \cite[BP;][]{Titov&al1993}  along the photospheric PIL or by  quasi-separatrix layer (QSL), whose core is a hyperbolic flux tube \cite[HFT][]{Titov02} that displays an X-shape when viewed along the PIL (Section\,\ref{ss:MFR}). 
An example of an HFT is given
in Panel (a) of Figure \ref{fig91}. This Figure 
contains maps of the  force-free parameter $\alpha$
($\propto$ the parallel electric current), and 
as shown by \citet{aulanier2005}, coronal currents maps trace QSLs. 
\textbf{Pure SMAs possess neither of these topological features}. 

Instead, they exhibit a gradual continuous transition from the sheared core to the current-free overlying arcades. This ``coronal magnetic shear gradient'' is a direct consequence of the prescribed smooth shearing profile at the solar surface (see Figure\,\ref{fig7}).
We note that, under non-ideal MHD conditions, \textsl{transient\/}, low coronal dips could potentially appear in SMAs, during SMA merging. These dips occur along short PIL sections and do not reach all the way to the photosphere, so they will not appear as "traditional" BPs \citep[][]{aulanier2006}. They are expected to be rare. However, with the increased availability of chromospheric vector magnetic field measurements, it is possible 
they could be observed
as 'chromospheric' BPs. We believe that the investigation of photospheric versus chromospheric BPs in high resolution magnetograms in multiple atmospheric layers could be a fruitful discriminator between MFRs and SMAs in the near future.

\subsubsection{Observational Signatures of Eruptions from SMAs}
We now turn to observational interpretations in favor of SMAs. Two models are mainly employed to explain how an SMA can erupt. The tether-cutting model \citep[e.g.][]{Moore&Roumeliotis1992, moore2001, 2016SoPh..291.2017P} is based on  the assumption that magnetic reconnection between two loop-like sheared arcades within the AR core field creates an MFR and drives the eruption. The second approach, known as the ``breakout model'' \citep{antiochos1999}, postulates that shearing motions drive the expansion of a core SMA, which may erupt only after it breaks through the overlying restraining fields via  reconnection at a magnetic null point. 

\noindent\textit{'Tether-cutting' vs. Breakout:\/} There are important differences between the two models: (i) tether-cutting can occur in either a bipolar or multipolar configuration, whereas the breakout model requires a multipolar magnetic configuration, and (ii) both models predict pre-eruption brightenings but at different locations. For tether-cutting  the brightenings occur at the AR core  while for the breakout model they occur at remote locations (and may be fainter). These two major differences are often used to postulate the pre-eruption magnetic configuration. We note that breakout could also trigger the eruption of a pre-existing MFR but in such a case the timing between flare and breakout reconnection may much tighter than in the case of "standard" breakout.

\noindent\textit{Crinkles:\/} In a series of studies, Sterling et al. investigated the formation and distribution of transient small-scale brightenings observed in UV spectral lines away from the core of the AR, which they called crinkles. \cite{2001ApJ...560.1045S}, \cite{2001ApJ...561L.219S}, and \cite{2004ApJ...602.1024S} found that the crinkles occurred before the core brightenings. This timing is consistent with the breakout model, which advocates the eruption of an SMA. It can be argued, though, that a pre-existing MFR in a quadrupolar configuration may become unstable even without involving breakout reconnection \citep[e.g.,][]{2007ApJ...671L.189A}, and that such an eruption may produce crinkles in a similar way. Also, the eruption of a pre-existing MFR may be triggered by breakout. 

\cite{2016ApJ...820L..37C} presented a rare case, where AIA 94\AA\ loops are seen to form a nearly ideal quadrupolar breakout configuration, followed by a CME and an X-class long-duration flare. While the external reconnection events (crinkles) do suggest that initiation of an eruption may take place without involving internal reconnection, they, however, do not allow to unambiguously discriminate between eruption of an MFR and that of an SMA. Therefore, we suggest that the timing and location of EUV brightenings may say something about the trigger mechanism, but possibly nothing about the nature of the pre-eruptive configuration.

Reconnection between erupting and surrounding fields may complicate the interpretation. \cite{2001JGR...10625227S} reported an event that obviously started from internal reconnection followed by the ejection of the core field, which later reconnected with external overlying fields, leading to the appearance of crinkles. \cite{2017ApJ...845...26J} analyzed a ``three-ribbon’’ flare and concluded that first an MFR formed and erupted following the tether-cutting scenario, and then the erupting fields reconnected with large-scale enveloping fields, which led to signatures similar to those expected from the breakout mechanism. 
If this scenario is common it will make it difficult to say whether breakout reconnection was the cause or just a result of an eruption, unless there is a clear time difference between the appearance of crinkles and flare loops.

\subsection{Magnetic Flux Ropes (MFRs)}
\label{ss:MFR}

In Section~2, we discussed three mechanisms that can generate MFRs; namely, photospheric flux cancellation, reconnection in the lower solar atmosphere 
between pre-existing and emerging flux or between pre-existing flux systems, and bodily emergence of an MFR. We now review the theoretical expectations (Sec~\ref{sss:MFR_theory}) and observational signatures (Sec.~\ref{sss:MFR_obs}) suggested so far in the literature.

\subsubsection{Theoretical Expectations} \label{sss:MFR_theory}

\begin{figure*}[h]
\centering
\includegraphics[width=.57\linewidth]{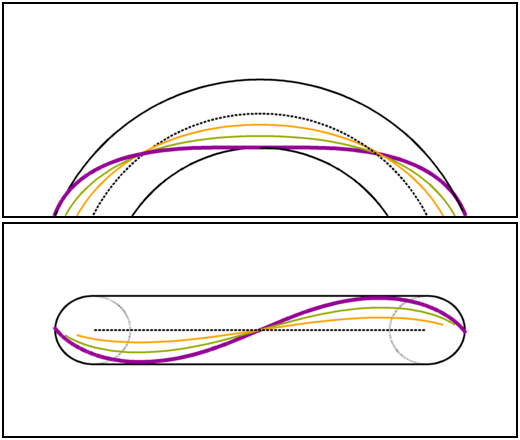}
\caption{Side and top view of an MFR of the minimum twist number that just yields a dip of a field line in the MFR. This tends to be the field line at the bottom edge under the apex (shown in dark violet) for a symmetric MFR whose twist profile is not strongly peaked at the axis. The minimum twist number is close to unity, $N \approx 1$, and only weakly dependent upon the geometry of the MFR \citep{Priest&al1989}.}
\label{fig9}
\end{figure*}

Since magnetic twist is a major element of MFRs, it is important to consider the expected number of turns in pre-eruptive MFRs. This can be derived from a number of theoretical and modeling considerations. First, since an MFR arches upward from its photospheric footprints, it will contain dipped field lines only if its twist reaches a (weakly geometry-dependent) minimum value, unless the MFR has a straight horizontal section. The schematic in Figure~\ref{fig9} illustrates that dips tend to form first at the bottom edge of the rope under the apex and require about one field-line turn if the rope arches upward (see \citealt{Priest&al1989} for the detailed geometric dependence).

Second, a twist slightly exceeding one turn typically results when the inner part of an SMA is transformed into an MFR by reconnection under the rope axis. Since the reconnected flux passes over both legs of the arched magnetic axis and under the axis at or near its apex, it makes exactly one turn between the points where it passes over the axis; the remaining sections of the reconnected flux, down to its footprints, add a relatively small fraction of one turn to the total twist (see the schematic in Figure~\ref{fig9}). The formation or enhancement of MFRs by flux cancellation \citep{vanballegooijen89,aulanier10,green11,zuccarello15}, during eruptions \citep{lin2004, Patsourakos&al2013}, and even in many flux-emergence simulations \citep{manchester2004,archontis2012, leake13a,Syntelis_etal2017} involves such reconnection, so that MFRs associated with FCs and eruptions are likely to possess a twist slightly above one turn.

Third, arched MFRs without field-line dips, i.e., of less than one field-line turn, visually resemble an SMA. Such ropes are unlikely to be inferred in images of coronal structures that outline field lines, i.e., from prominence threads or EUV/SXR sigmoids. 

Fourth, NLFFF models for FCs and their environment in the corona can be obtained through extrapolation of a vector magnetogram \cite[e.g.,][]{Wiegelmann2008}, magneto-frictional relaxation \citep[e.g.,][]{vanballegooijen00}, through the flux rope insertion method \citep{vanballegooijen04}, as well as through evolutionary methods \citep{Yeates&Mackay2009, mackay2011, yardley2018b, chintzoglou2019, price2019}. The cross-section average of the inferred twist ranges from below unity to $N\approx\!2$. Using extrapolation, an MFR was found in a number of cases \cite[e.g.,][]{Canou&al2009, YellesChaouche&al2012, GuoY&al2013,cheng2014,chintzoglou15,WangH&al2015,LiuR&al2016, ZhaoJ&al2016, Green&al2017, james2018,mitra2018,DuanA&al2019,woods2020}, while an SMA resulted in others \cite[e.g.,][]{SunX&al2012}. 

For some ARs, one extrapolation technique found an MFR and others found an SMA \citep{Schrijver&al2008, Inoue&al2016, JiangC&al2016, Amari&al2018}. The experience with this technique over many years indicates that it often struggles to find an MFR \cite[e.g.,][]{Schrijver&al2008} and that success in finding one is more reliable than failure. Success in finding an MFR is corroborated by the presence of bald patches in the filament channel in most cases. 

Fifth, the excitation of the helical kink instability, a possible mechanism of eruption onset, requires a twist of at least $N>1.25$ in the current-carrying cross section of the MFR \citep{Hood&Priest1981}, but more likely $N \gtrsim 1.5\mbox{--}1.75$ in realistic configurations with a guide field component \citep{torok04, Fan_etal2004}. Therefore, pre-eruptive MFRs with high 
twist ($N>2$), if they exist, should not be stable for long intervals, except under certain conditions such as a strong guide field or a very thin MFR \citep[see the discussion in][]{torok14b}. This suggests that most pre-eruptive, stable MFRs should be weakly to moderately twisted, within the range of $N\approx1\mbox{--}2$ turns. 

Another important aspect is the ``visibility'' of the twist and, related to this, the radial twist profile. An MFR can have \textit{any} radial twist profile, $N(r)$. Cases with $N(r)$ peaking at the axis ($r \to 0$) have been obtained in simulations of flux emergence \citep[e.g., Fig.~3b in][]{torok14a}. So-called hollow-core MFRs (see also Section\,\ref{sss:SMA_pre}), whose twist profile peaks at the surface ($r \to a$), have been found to yield good models of FCs in decaying ARs which are dominated by flux cancellation (e.g., \citealt{bobra2008, savcheva2009, SuYN&al2011}; Figure~\ref{fig91}). Especially the latter models have received widespread attention in the modeling of FCs. They consist of highly sheared but (possibly very) weakly twisted flux in the core, surrounded by a layer of clearly (but not necessarily highly) twisted flux at the surface of the structure. As a whole, the structure is an MFR, although the non-potentiality of the core flux is strongly dominated by shear. The field lines in the core flux obviously display very small twist \citep[e.g.,][]{savcheva2009}, and the same can be expected of field line tracers in solar observations. However, the latter is a difficulty for a much wider range of MFRs, except those particularly highly twisted around the axis. Any twist near the MFR axis is much less visible than twist of the same magnitude near the surface. See illustrations of this phenomenon for a uniformly twisted MFR, e.g., in \citeauthor{Priest2014} (\citeyear{Priest2014}; Figure~3.5) and \citeauthor{torok10} (\citeyear{torok10}; Figure 2).

\subsubsection{Observational Signatures} \label{sss:MFR_obs}

Given that several of the proposed MFR tracers apply uniquely to either QS or AR filament channels, we discuss them separately below. We attempt to draw parallels between them, where applicable.    

\noindent
\textbf{MFR Signatures in QS Filament Channels}
\\
The model for a filament channel hosting a quiescent prominence is given by an MFR in many studies  \citep[][see Section~3.1 for SMA models]{Kuperus&Raadu1974, vanballegooijen89, Priest&al1989, Low&Hundhausen1995, Aulanier&Demoulin1998}. This is based, first of all, on the observation that most quiescent prominences exhibit inverse polarity \citep{Leroy&al1983,leroy1984,Bommier&al1994,Hanaoka&Sakurai2017}. 
Inverse polarity results naturally if the prominence material is trapped in the magnetic dips in the lower half of an MFR, but requires special conditions if the prominence material is trapped in an SMA (see below). 

\begin{figure*}[h]
\centering
\includegraphics[scale=0.3]{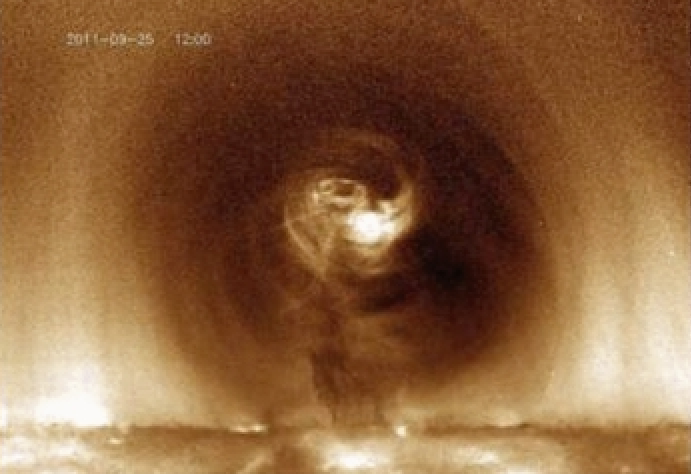}
\caption{Quiescent prominence and its coronal cavity observed by AIA in the 193~{\AA} channel \cite[from][]{LiX&al2012}. \copyright AAS. Reproduced with permission.
}
\label{fig10}
\end{figure*}

\textit{Magnetic Field Dips:\/} Inverse polarity dips are possible in an SMA if its end sections lean over the PIL and bulge upward as discussed in Section\,\ref{sss:SMA_pre}. This happens if the main flux sources on either side of the PIL are considerably displaced along the PIL, which is possible for a strongly sheared region. However, for the effect to be significant, the displacement must be such that the middle section of symmetric and strong confinement must be relatively short compared to the length of the SMA. Otherwise, the SMA field lines will be closely aligned with the PIL in the middle section, so that the inverse field component will be much weaker than the axial component, i.e., it will not reach the observed angle of 10--30 degrees from the PIL \citep[e.g.,][]{Hanaoka&Sakurai2017}. While the displacement of the main flux sources may satisfy this requirement for some ARs and intermediate filament channels, it seems unlikely to be realized for the filament channels of long quiescent prominences, which can reach a solar radius in length or more. This obviously contrasts the arguments in Sec.~\ref{sss:SMA_pre} regarding the ability of SMA models to contain IP dips. 
\textbf{Detailed, quantitative observations of shear patterns in the photosphere are required to test these assertions.}

\textit{Thread Orientation:\/} The typical orientation of the threads relative to the axial field of quiescent filaments, combined with the assumptions of force-free equilibrium and the presence of a net current along the filament, is consistent with an MFR or the MFR component \textit{in a hybrid structure only\/}. This results from the following reasoning. The field lines in a right-handed SMA and in the top half of a right-handed MFR point to the right relative to the direction of the axial field of the prominence in vertical projection, but point to the left in the bottom half of the MFR. The force-free equilibrium of the current-carrying field of the filament requires that any net current must be directed such that its Lorentz force with the so-called strapping field (i.e., the field due to external sources passing over the filament), $\boldmath{J \times B_\mathrm{ext}}$, points downward. Therefore, the current points into the direction of the axial field in so-called sinistral filaments \citep[as defined in][]{martin1998}, which are, therefore, right-handed. The threads in sinistral filaments point to the left, which is also the prevailing direction of deviation of the filament barbs from the axial field and spine of the filament (see, e.g., Figures 10 and 11 in \citealt{martin1998}). Therefore, the direction of the threads (and barbs) in quiescent filaments
demonstrates that their structure is that of an MFR.
The mirror image, with the opposite handedness and 
opposite
direction of threads and barbs, is observed for so-called dextral quiescent filaments, which, of course, equally clearly indicate an MFR field. 
Finally, since FCs embody high magnetic shear at the PIL, the  flux complex they are embedded in generally carries a net current \citep{torok03,torok14a,dalmasse15}, which is also observed \citep{LiuY&al2017,Kontogiannis&al2017}.

\textit{Cavities:\/} 
The low-density cavities of quiescent prominences \citep[for a review on quiescent cavities, see,][]{gibson2015}
provide strong evidence for the evolution from SMA to MFR structure via resistive processes as discussed in Section 2.2. Case studies of high-lying (i.e., old) prominences revealed elliptical cavity shapes fully detached from the limb, suggestive of an MFR viewed along the axis and incompatible with an SMA. The prominence material is located in the bottom half of the ellipse and sometimes shows concave-upward fine structure \citep{Gibson&al2010,Karna&al2015b}. A particularly clear example is shown in Figure~\ref{fig10}. \citet{Karna&al2015a} studied over 400 cases, focusing on the properties in the middle of the cavity's life time. The overall 3D topology of cavities could be characterized as a long tube with an elliptical cross section. Most cavities then were partial ellipses, whose midpoint was elevated above the limb, with the prominence material reaching up to about the midpoint, i.e., being situated in the lower half of an apparent full ellipse centered on the observed midpoint and covering the prominence material, except for its feet. 
\textsl{This indicates the SMA-to-MFR transformation theoretically expected from the flux cancellation process, with the area of the cavity outside the virtual ellipse still being an SMA\/}. The geometry also indicates magnetic dips at the position of the prominence material. \citet{Forland&al2013} showed that partially elliptical cavity shapes typically evolve into full ellipses at gradually increasing height, which eventually fully detach from the limb and then develop an inverse teardrop shape. They also found that prominences in cavities at the latter stage are about seven times more likely to erupt than those in elliptical and semi-circularly shaped cavities and suggested a combination of 
current sheet formation at the bottom tip of the cavity and MFR instability as the cause for the increased propensity for eruption. Slow reconnection in the CS would supply the cavity/MFR with flux, such that its center rises, eventually reaching the critical height for onset of the torus instability \citep[][]{kliem06}.
These findings convincingly suggest that the SMA-to-MFR evolution continues throughout the lifetime of quiescent prominences and directly leads to their eruption.
\textbf{We conclude that the magnetic fields carrying quiescent prominences are in a transitional SMA-to-MFR stage throughout their lifetime.}
Therefore, the notion that FCs carrying quiescent prominences are either SMAs or MFRs appears to be overly restrictive.
Finally note, that sometimes eruptive filaments/prominences
show evidence of helical structure \citep[e.g.,][]{vrsnak1990}.

\textit{Spinning Motions:\/} 
Using high-resolution, high-cadence coronal images obtained from STEREO/EUVI, \citet{WangY-M&Stenborg2010} found that some coronal cavities appear as continuously spinning structures, with sky-plane projected flow speeds in the range of 5--10~km\,s$^{-1}$ that often persist for several days. They argued that such persistent swirling motions provide strong evidence that the cavities are MFRs viewed along their axis.
SMA hybrids with helical field lines (Fig.~\ref{fig8}) 
could exhibit swirling motions in their IP/NP interface(see dips discussion in \ref{sss:SMA_pre}). In this case, however, the center of the swirling motions will be located \textit{below} the prominence apex. In the MFR case, the center of the swirling motions lies above the prominence top, consistent with the existing observations.
Moreover, helical
field lines exist only in a shell of relatively small flux content, whereas the main central part of the flux in this hybrid structure is only sheared, not helical. 

However, we should be conscious of the impact of projection effects on observational studies. \citet{Bak-Steslicka&al2013} using COMP observations, found concentric rings of the line-of-sight velocity within cavities; the velocity rings provide strong evidence of an MFR structure, as they suggest flows along flux surfaces of an MFR. On the other hand, \citet{schmieder2017} reconstructed a seemingly helical prominence observed by IRIS with slit-jaw images and spectra and deduced that the spiral-like structure of the prominence observed in the plane of the sky is mainly due to the projection effect of long arches of threads. In other words, the actual 3D morphology of the cavities needs to be taken into account for the proper interpretation of prominence fine structure as discussed in \citet[e.g.,][]{gunar2018}.

\textit{Polarization:\/} Synthetic observations of the linear polarization in forbidden IR lines in the corona show little differences between cylindrical MFRs and sheared arcades \citep[][]{Rachmeler&al2013}, unless observations at altitudes below the inner field of view (FOV) of COMP are available. On the other hand, \citet{Rachmeler&al2013} showed that \textbf{observations of the circular polarization can distinguish between SMAs and MFRs, since in the former case the strong signal is close to the limb whereas in the latter case a circular ring of enhanced signal is anticipated} \citep[see Figures 5 and 7 in][]{Rachmeler&al2013}.\\

\noindent
\textbf{MFR Signatures in AR Filament Channels}

\textit{Thread Orientation:\/} Measurements of the magnetic field in AR prominences are not yet possible, because these form at low heights, where scattered light from the limb is too strong. For AR filaments well within the solar disk, vector magnetograms have revealed inverse field direction across the photospheric PIL, i.e.\ BPs, for some FCs \citep{JiangC&al2014,yardley2016,ChandraR&al2017,Kliem&al2020}. Measuring the direction of threads in AR filaments requires high resolution and is still in its early stages. The data available so far tend to point  to a direction opposite to that of quiescent filaments (NP instead of IP). A clear example is shown in \citet{WangH&al2015}, where the threads of a filament contained in left-handed field point to the left, a result consistent in principle with both an MFR and SMA. On the other hand, a few AR filaments have been found to show winding threads directly indicating an MFR;  see an example suggestive of one field-line turn in \citet{XueZ&al2016}.

\textit{Cavities:\/} Cavities around AR filaments are often seen to form during eruptions as an element of the three-part structure of white-light CMEs \citep{Illing&Hundhausen1985}, but, with one exception, have not yet been observed as equilibrium structures prior to eruptions \citep{patsourakos2010a}. This is because AR filaments are very low-lying, so that their cavities are likely obscured by the AR emission structures. An AR cavity has been observed, in one case, for about one hour during the slow-rise phase prior to the onset of a fast CME. The helical fine structure of the embedded filament was suggestive of an MFR \citep{ChenB&al2014}. 

\textit{Dips:\/} From the above observations, in particular from the structure of AR prominences, it is clear that prominence material in ARs is far less associated with magnetic dips than in the QS. This is not yet fully understood but is consistent with the following considerations. Since the nearly horizontal fields in AR filament channels have much  lower altitudes, they can be supplied with cool material from below (e.g. via chromospheric jets) frequently and thus the material of each individual thread does not need to be trapped for a long time for a continuous presence of the prominence as a whole. This is in line with numerical simulations of prominence support by long field lines without dips \citep{karpen2001, luna2012}. Magnetic dips are much more difficult to create by the prominence material, because the Lorentz force in ARs exceeds the gravitational force much more than in the QS. Indeed, AR prominences are much more variable and short-lived than their quiescent analogs, with up and downflows being omnipresent. Consequently, their threads are more likely to populate both concave-upward and concave-downward field lines, so that an MFR structure is more difficult to infer observationally. However, ARs provide further opportunities to infer MFRs, which we discuss below.  

\begin{figure*}[h]
\centering
\includegraphics[scale=0.3]{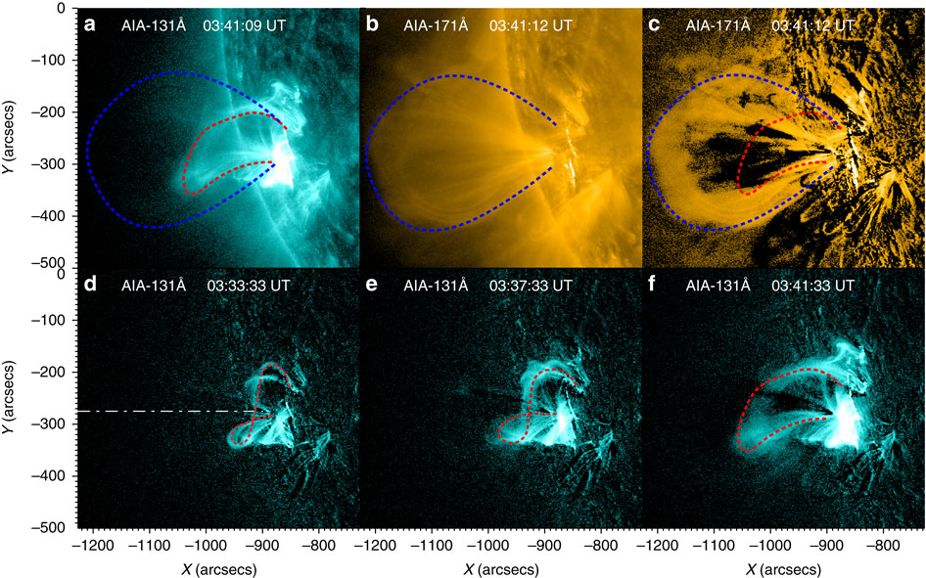}
\caption
{Evolution of a hot channel (presumably an MFR) during the March 8, 2011 eruption. The channel (marked by the red dotted line) is visible in the hot AIA 131~{\AA} passband (a), but absent in the cool 171~{\AA} passband (b). (c) Difference image (171~{\AA}; base image at 03:20:41~UT), showing the compression front of the eruption (marked by the blue dotted line). (d)--(f) Sequence of difference images (131~{\AA}; base image at 03:20:09~UT), showing the evolution of the hot channel (red dotted lines). The channel apparently transformed from a writhed sigmoidal shape into a semi-circular shape \cite[adopted from][]{ZhangJ&al2012}.
}
\label{fig12}
\end{figure*}

\textit{Sigmoids:\/} The coronal signature of many AR filament channels is a SXR sigmoid. As  discussed in Section \ref{sss:cance_obs}, single-S sigmoids are considered to be a strong observational indication of an MFR, especially when their middle section crosses the PIL in the inverse direction \citep{Green&Kliem2009,green11} and when their ends turn around strongly to point toward the center of the sigmoid \citep{Kliem&al2020}. Diffuse sigmoids, on the other hand, are composed of many loops of a regular arc or J shape, which only collectively form an S shape. Therefore, they are very suggestive of an SMA structure. It is very likely that the transition from the diffuse form via double Js to single S sigmoids, 
which occurs during extended periods of photospheric flux cancellation \citep{Green&Kliem2009, green11, savcheva2012}, outlines the SMA-to-MFR transformation conjectured by \citet{vanballegooijen89}, \citet{moore2001}. The transformation relies on persistent, ubiquitous flux cancellation events 
(one of which is sketched in Figure~\ref{fig6}), 
where reconnection progressively increases the highly sheared and
typically only
moderately twisted flux of the FC. However, the twist
may or may not increase in this process; compare the schematics
in \citet{vanballegooijen89} and Green et al (2011).

\textit{Hot channels:\/} A new line of observational evidence for MFRs emerged with the so-called EUV ``hot-channel'' structures revealed by \textsl{SDO}/AIA observations \citep{ZhangJ&al2012, ChengX&al2013,Patsourakos&al2013, Nindos&al2015,ZhouZ&al2017,veronig2018}. These structures exhibit a sigmoidal shape when seen against the disk \citep{LiuR&al2010} and resemble a hot blob or arc when seen above the limb \citep{ChengX&al2011,ChengX&al2014}. 
 
The hot channel can be clearly seen in the AIA 131~{\AA} passband, which is sensitive to high plasma temperatures $\sim$10~MK, while it is completely absent in 
the AIA 193~{\AA} passband, which is sensitive to AR 
coronal temperatures ($\sim$ 2 MK), and in the  AIA 171~{\AA} passband that is sensitive to quiet corona/transition region temperatures of $\sim$0.6~MK (Figure~\ref{fig12}). 

Since these hot temperatures are higher than those of SXR sigmoids, the channels are more closely associated in time with flaring activity, which provides plasma heating through reconnection.
The statistical study of \citet{Nindos&al2015}  showed that 49 $\%$ of major eruptive flares (above GOES M-class) involve hot channels.

These hot channels frequently exhibit hook-shaped ribbons during flares, an indirect indication for an MFR \citep[][]{demoulin1996, cheng2016, ZhaoJ&al2016}. Note that the hooks would extend into spirals for highly twisted MFRs \citep{demoulin1996}.
In the case of a pre-existing MFR, hook-shaped ribbons are formed right from the flare onset.
MFRs possess topological features, such as HFT/QSLs and BP/BPSS (see also section 3.1.2). Current sheets are prone to form along these topological features, typically on Alfv\'{e}nic timescales, when the magnetic field is evolving. As a result, flare-reconnection can start early during MFR eruptions and illuminate ribbons, both under the flare CS and the hook-shaped ribbon extensions \citep[e.g.,][]{aulanier10,zuccarello15}. In the case of an SMA though, there could be some delay between flare onset and their 
appearance. This is because quasi-linear ribbons will appear first under the flare CS, and then their hook-shaped extensions will develop once the MFR starts to form during the eruption. Therefore, observations of hook-shaped ribbons have a limited potential to distinguish between SMAs and MFRs, unless possibly when observing them during  confined flares preceding an eruption.

Unlike filaments, which often become fragmented during the eruption, the EUV hot-channel structures maintain their coherence throughout the eruption process.
Unlike SXR sigmoids, which quickly fade away into the background (thus cannot be traced for long), the channels can often be continuously traced until they leave the AIA FOV. 

Most of the identifications and studies of hot channels have focused on occurrences in the course of eruptions. 
Here, we are interested in evidence for pre-existing MFRs, meaning MFRs that formed 
clearly before the eruptions, so we focus on such evidence below. In events that feature a hot channel already in the slow-rise phase, the channel displays a morphology similar to its appearance after the main eruption onset \citep[][]{ZhangJ&al2012,ChengX&al2013}. 
\citet{Patsourakos&al2013} described a case where an MFR was formed long before an eruption. They identified a hot structure that formed during a confined flare and  erupted seven hours later. 
This hot structure had an oval (blob-like) shape and was formed above a cusp straddling the corresponding post-flare loop arcade. The cusp  then corresponded to the CS associated with the ascent of the magnetic flux which gave rise to the corresponding confined flare. MFR formation (or enhancement) during eruptive flares, is a common element of MHD models, whether they do \citep[e.g.,][]{vanballegooijen89} or do not include \citep[e.g.,][]{karpen2012} a pre-eruption flux rope. Another important take-away from  \citet{Patsourakos&al2013} was that, by limiting the observational window only around the eruptive flare, one would have missed the confined flare event and hence would have been led to different conclusions about the formation time of the erupting MFR. \citet[][]{chintzoglou15}, \citet[][]{kumar2017}, \cite{james2018}, \citet{liu2018}, and \citet{Kliem&al2020} essentially reached the same conclusion for other events: MFRs were formed during confined flares. NLFFF extrapolations revealed twisted field lines in the places where hot  structures were formed during some of the reported confined flares. 

\begin{figure*}[h]
\centering
\includegraphics[scale=0.3]{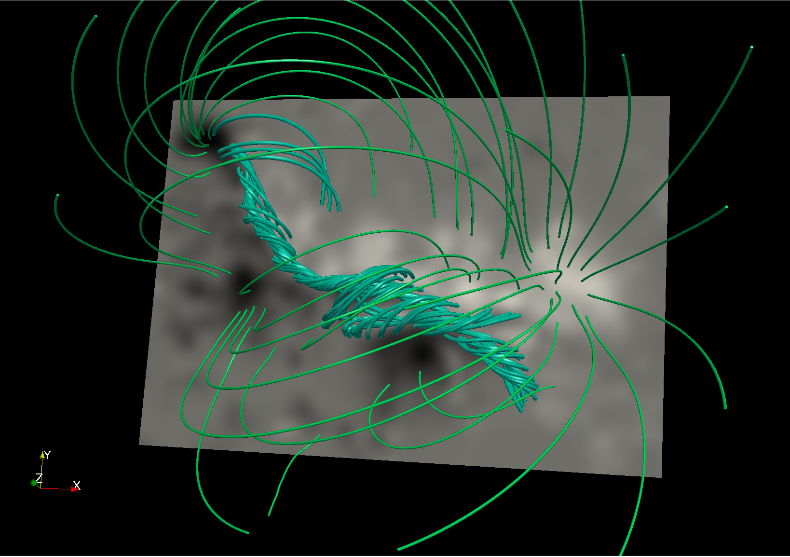}
\caption{NLFFF extrapolation of the magnetic field of a filament channel, showing an incoherent MFR structure (two MFRs separated by a short SMA section) shortly before its two relatively coherent MFR sections successively erupted to cause two major CMEs \citep[from][]{chintzoglou15}. \copyright AAS. Reproduced with permission.
}
\label{fig121}
\end{figure*}

Our review of the SMA and MFR signatures can be summarized as follows. MFRs in FCs are expected to form through a transition from an SMA, 
often driven by flux cancellation. Since flux cancellation operates gradually and intermittently due to the fragmented structure of the photospheric magnetic flux, FCs should be in a transition state for a long time. Moreover, as long as flux cancellation episodes still occur, an already existing MFR will be enhanced by additional flux. The additional flux does not necessarily extend along the whole FC, especially not for long ones, and will branch off from the main body of the MFR in an arcade-like manner toward its photospheric roots, which may be seen as barbs. Therefore, although the observations cited in Section~\ref{sss:MFR_obs} clearly indicate that (at least) quiescent prominences form when the transition to an MFR structure has strongly advanced, the MFR may still be completing its transition or enhancement. One should not expect an MFR that is perfectly coherent from end to end. During the transition, one may find a mix of arcade-like and rope-like sections(both
(typically in a small number) with a growing degree of coherence.
These may take the form of flux branching off from an otherwise fully developed MFR \citep{GuoY&al2013} or of a section of SMA structure adjacent to an MFR \citep{GuoY&al2010} or between two MFRs  \citep[][see Figure~\ref{fig121}]{chintzoglou15} 
or
coexisting sheared and twisted field lines of different extension (Figure 3(e)).
It is an open question how coherent an MFR has to be 
(in terms of flux connecting its two main footpoints)
to allow the initiation of a CME by a loss of equilibrium or, equivalently, an ideal MHD instability. In addition to this gradual process of MFR formation, the observations of EUV hot channels indicate that MFRs can be formed or extended by reconnection on a much shorter time scale during periods of low-level flaring activity 
which precede a subsequent major eruption into a CME. These can be a confined flare or the slow-rise phase leading to the major eruption \cite[e.g.,][]{LiuR&al2010,ZhangJ&al2012,Patsourakos&al2013}.

\subsection{Differences in eruption signatures between SMA and MFR} \label{sss:SMA_eruptive}

Although our focus is on the pre-eruptive magnetic field configuration, we assess briefly below some potential SMA/MFR differences during the eruption phase as these are often used in the literature.

\textit{Flare Timing:\/} A sufficiently sheared SMA will collapse to form a vertical CS extending upward from the PIL. Reconnection in the CS tends to set in immediately, which marks the onset of a flare, and forms an MFR at the tip of the CS \citep[e.g.][]{Mikic&Linker1994,karpen2012}. Depending on parameters, the MFR can start to rise immediately or could be delayed. Hence, in an erupting SMA, a flare will start prior to or simultaneously with the onset of an eruption. In comparison, the timing between flare and CME onset is reversed in an MFR if the MFR is of BP topology. In this case, a vertical CS forms only when the unstable upper part of the MFR has risen sufficiently \citep{gibson06}. However, if an X-type HFT exists at the underside of an unstable MFR, then the HFT collapses into a vertical CS, and reconnection commences, simultaneously with the onset of the rise, i.e., flare and CME commence simultaneously. 
However, since flare onset and CME onset are rather well synchronized in the majority of eruptions \citep{Zhang&Dere2006, temmer2010,ChengX&al2020},
with a  trend though that CME rise starts before the impulsive flare energy release (in terms of the flare SXR and HXR emission, respectively) as shown in \citet[][]{bein2012, stoiser2012}, the time difference between them 
can only rarely be used to discriminate between SMA and MFR as the pre-eruptive structure.

\textit{Partial Eruptions:\/} It is frequently observed that the top part of a filament channel rises and erupts explosively, while the bottom part remains on the Sun, apparently undisturbed \citep[e.g.,][]{pevtsov02}. Partial eruptions from an SMA configuration seem straightforward to understand as flare reconnection and CS formation can occur anywhere along such configurations transforming only part of the SMA into the escaping MFR \citep[e.g.,][]{cheng2014}.
H$\alpha$ observations sometimes show flare ribbons appear and spread far from the observed dark filament, implying that the eruption and reconnection involve only the FC magnetic flux \textit{above\/} the flux carrying the cool material. 
However, MHD simulations have shown that a single MFR can split via internal reconnection in the early phase of its eruption \citep[e.g.,][]{gibson06,kliem14b}, as previously suggested by \cite{gilbert01}. Such a split will occur when the main part of the MFR, including its axis, satisfies the criterion for an instability while the bottom part of the MFR is tied to the photosphere in a BP. Other works have invoked the existence of a so-called ``double-decker'' configuration of two MFRs lying on top of each other prior to an eruption and assuming that only the upper one, which may not carry filament material, erupts \citep[e.g.,][]{liu.r12,cheng2014,kliem14b,dhakal2018}. It seems, therefore, that while the vertical splitting of the FC flux follows naturally for  SMA topology, it requires special conditions to explain with an MFR topology.

\textit{Ideal Instabilities:\/} 
The internal current distributions  differ strongly in SMAs and MFRs, which is relevant for the development of ideal instabilities that may initiate an eruption. SMAs and MFRs are both force-free, so they contain field-aligned currents whose paths follow exactly the associated stressed field lines. Just like the field-line shapes and connectivity, current paths are very different in SMAs and MFRs. In particular, the currents in an MFR go all the way from one end of the rope to the other. This is shown, e.g, in Figures \ref{fig9} and \ref{fig13}. In an SMA, on the other hand, magnetic shear decreases with distance from the PIL. Therefore, electric currents can connect areas of very large and very small shear, as can be seen by looking at the field lines in, e.g., Fig.\,\ref{fig8}.

Such a coronal distribution of electric currents is unfavorable for the occurrence of ideal loss-of-equilibrium scenarios and/or the torus instability. This is because, neutralized currents shield the main currents from which Laplace forces are calculated, so there can be no repelling between distant-components and/or subphotospheric-images of the modeled currents "wires".
Along the same lines, the differential shear by itself cannot produce local twists of more than about half a turn in an SMA. Therefore, SMAs are not prone to the helical kink instability,  either. It follows that an isolated bipolar SMA is an ideally stable configuration, so that it cannot erupt. This was first demonstrated by \citet{Roumeliotis&al1994}, \citet{Mikic&Linker1994}, and \citet{Amari&al1996}. The only way to make an SMA erupt is to provide an external perturbation, such as, e.g., in the magnetic breakout model, and to produce a CS in which flare reconnection can occur. 
In other words, the detection and quantification of an ideal instability as the \textit{driver} of an eruption appears to be the clearest discriminator between an SMA and an MFR as the eruption origin. 
Unfortunately, the potential of ideal instabilities is (yet) to be fully investigated, mostly due to observational constraints. For example, it is difficult to quantify the role of the torus instability because incomplete coverage of the photospheric magnetic field and lack of high-cadence multi-viewpoint EUV imaging complicates the inference of the height of the eruption onset and the magnetic field decay index at that height \citep[see detail, e.g., in][]{ChengX&al2020}. 

Additionally, in highly energetic cases, an SMA could transform into an MFR  too rapidly for our observational cadence to provide evidence
for an instability prior to the inevitable onset of reconnection.

\section{Summary}
\label{s:recap}

The goal of our ISSI team was to investigate and to resolve, if possible, the debate on whether the pre-eruptive magnetic configurations of CMEs are either SMAs or MFRs. 
Our objectives were to: (1) examine the nature of pre-eruptive configurations of CMEs as revealed by current observations, numerical modeling, and theoretical inferences, (2) identify the origins (theoretical and observational) of the debate and then (3) attempt to reconcile the differences and sketch a path forward on this issue.

To this end, we reviewed the current state of knowledge on the formation mechanisms and magnetic configurations of the pre-eruptive structure of CMEs (Sec.~\ref{s:FC-formation}). We then reviewed the theoretical expectations and observational signatures that are commonly used to interpret observed features as SMAs or MFRs (Sec.~\ref{s:configurations}). 
In Table~\ref{t:table2}, we summarize those signatures and their association with either an SMA or an MFR. It is obvious that several signatures can be attributed to both types of pre-eruptive configurations.
The main reasons for the dearth of unique signatures are, on one hand, the inherent ambiguities in imaging and photospheric magnetic field observations, which form the bulk of the observables, (e.g., line-of-sight integration, signal-to-noise ratios, spatial resolution, lack of multi-height magnetic field measurements, non-force free photospheric magnetic field,
etc.)  and, on the other hand the fact that SMAs and MFRs in FCs often exist in states with partly similar properties (e.g., nearly straight field lines in the center of both highly sheared SMAs and highly sheared but weakly twisted MFRs); namely, the co-exist in a \textit{hybrid} state.

\begin{table}[!ht]
\begin{center}
\caption{Observational signatures of pre-eruptive SMAs and MFRs. 
The term SMA refers primarily 
to the 3D strongly sheared S-shaped SMAs discussed in  Sec. \ref{ss:SMA}. IP and NP
refer to inverse and normal prominence polarity (Sec.
\ref{ss:SMA} and  \ref{ss:MFR}). BP and
HFT correspond to bald patches and hyperbolic flux tubes
(Sec. \ref{sss:SMA_pre}). 
 In the case of  SMA-MFR 
hybrids
the listed signatures refer to the flux that embodies the
SMA and MFR components of the hybrid structure. } 
\label{t:table2}
\begin{tabular}{p{2.2cm}p{0.8cm}p{0.8cm}p{6.6cm}}
\hline
\rowcolor{Gray}
\textbf{Observable} & \textbf{SMA} & \textbf{MFR}  & \textbf{Remarks}
\\ 
\hline
single-S sigmoid   (\ref{sss:cance_obs},\ref{sss:MFR_obs})           &      N      &  Y   &  \\ 
diffuse sigmoid  (\ref{sss:cance_mod}, \ref{sss:MFR_obs})  &     Y       &  N    & \\
prominence dips & Y & Y & SMA: IP and  NP dips  
in prominence main body  and top/periphery respectively
for strong shear and large departures from 2.5D (no dips for arched-shaped SMAs); MFR: IP dips (small-isolated patches of NP dips possible when PIL surrounded with parasitic polarities) \\ 
linear threads in prominences (\ref{sss:SMA_pre})    &  Y          &  Y   & 
 \\ 
elliptic or teardrop-shaped EUV cavities (\ref{sss:MFR_obs})   &      N      &  Y   &  \\ 
spinning motions in EUV cavities (\ref{sss:MFR_obs})   &  N           &  Y   & SMA: possible but only below prominence top (not observed)
\\ 
circular polarization in the corona (\ref{sss:MFR_obs}) & Y & Y 
 &  SMA: stronger near limb; MFR: circular ring above limb \\ 
BP (\ref{sss:SMA_pre})  &       N    & Y   & \\    
HFT (\ref{sss:SMA_pre}) &        N    &  Y   &  \\ 
flare-related EUV hot channel (\ref{sss:MFR_obs}) &    N        &  Y   &  
pertinent observations during confined flares 
prior to the eruption are better suited to probe
the pre-eruptive configuration
\\ 
hook-shaped ribbons  (\ref{sss:MFR_obs}) & Y          &  Y  & 
observed from flare onset for an MFR and possibly  
delayed for an SMA; 
pertinent observations during confined flares  before the eruption are
better suited to probe the pre-eruptive configuration\\ 
coronal currents (mapping only differential shear across the PIL) (\ref{sss:SMA_eruptive})   & Y  &  N &  \\   
coronal currents (end-to-end along the PIL) (\ref{sss:SMA_eruptive})   &     N       & Y    &   \\ 
CME acceleration at Alfv\'enic speeds (\ref{sss:SMA_eruptive}) 
&  Y   & Y        & SMA:  only at flare impulsive phase  onset; MFR: possible before flare impulsive phase onset.
\\  
\hline
\end{tabular}
\end{center}
\end{table}

We emphasize that the proper assessment of the pre-eruptive configuration of CMEs requires that the observables of Table~\ref{t:table2} are  applied to "truly" pre-eruptive states 
that exist up to the onset point defined in Section \ref{s:intro}, 
i.e., \textit{before} the speed of the ejected structure exceeds 100 km/s. It is prudent to search for these signatures from a few hours prior to the eruption up to the slow rise phase that often precedes CMEs \citep[e.g.,][]{zhang2004, Zhang&Dere2006}. A given signature could be visible only during particular times/conditions (e.g., the 'hot channels' observed in the AIA 131\AA \, passband discussed in Section \ref{ss:MFR}), and therefore, it is recommended to search for potential signatures at multiple instances during the pre-eruptive phase.

Finally, we were able to identify and reach consensus on several key issues that have fuelled the debate in the past:
\begin{itemize}
\item 
Observations and data-constrained modeling indicate that highly twisted pre-eruptive MFRs ($N>2$) are exceedingly rare. 
\item  The observational evidence for bodily emerging MFRs is scarce, which is in agreement with almost all flux-emergence simulations.
\item  
Different physical mechanisms (flux emergence/cancellation, helicity condensation, tether-cutting reconnection) can be involved in the formation of SMAs and MFRs, and their respective contributions depend on the evolutionary stage and location (AR, between ARs, QS) of the pre-eruptive configuration.
\item The amount of magnetic flux associated with SMAs and MFRs is hard to pin down either with modeling or observations. For example, magnetic cancellation studies suggest that a significant amount of the total AR flux can be available for inclusion into an MFR, while NLFFF modeling often suggests that MFRs contain only a small fraction of 
the AR flux. This is an area that requires attention (see next section).
\item In many cases, pre-eruptive configurations seem to undergo a slow transition from SMA to MFR during their evolution. This transition likely accelerates during the slow rise phase prior to an eruption.
\item The spatial coherence of 
the pre-eruptive magnetic configuration could be used to determine the stage of this transition, i.e., whether the configuration is closer to an SMA or an MFR state, and whether the configuration  is `ready to erupt'. Methods to quantify the coherence, using modeling or observations, should be pursued.
\end{itemize}

The most important conclusion from our deliberations is that the debate on SMA versus MFR may be  a 'red herring". Neither the observations nor the theory nor numerical modeling support a fixed interpretation of the pre-eruptive configuration as being either an SMA or an MFR. The most sensible conclusion is therefore, that {\bf the pre-eruptive state is a 'hybrid' configuration that contains both sheared and twisted field lines at varying proportions}. Moreover, one should not expect that the configuration is perfectly coherent from end to end.
In most cases, we are observing a magnetic system in transition from SMA to MFR. Depending on the state of this evolution, the configuration may be more ``SMA-like'' or more ``MFR-like''.

Therefore, the question now shifts to determining the degree of 'SMA-ness' or equivalently of 'MFR-ness' of the magnetic configuration of a given FC. Quantifying this degree may be performed by calculating
the (dimensional) ratio ${\mathcal{R}}_{\Phi_B}$ between the
sheared and twisted flux (see also
discussion of Section \ref{sss:defi}  on
hybrid structures). $\mathcal{R}_{\Phi_B}$ is a parameter
that may be derived from magnetic field cubes
resulting from either MHD simulations or from NLFF extrapolations
from magnetograph data. Another pertinent metric may be based on the
decomposition of magnetic helicity in self and mutual terms
(see section \ref{ss:condensation}). Clearly, new concepts
and tools are required to address 'SMA-ness' and 'MFR-ness'.  
Successful investigation of this question should lead to a better understanding of CME initiation and could eventually improve our ability to predict their occurrence by inferring  key eruptivity parameters pertinent to ideal MHD instabilities (e.g., twist, decay index of the overlying field) and magnetic reconnection (e.g., plasma heating and particle acceleration, reconnected flux). 
In the next section, we propose a path forward to make progress on the outstanding issues in the theory, observations, and modeling of pre-eruption FC magnetic structure.

\section{Path Forward} \label{s:path}
%
The final task of our ISSI team was to recommend a path forward for the difficult problems described in the previous sections. We compiled a list of recommendations for further studies; one set deals with observations and data analysis issues (Section\,\ref{ss:path_observations}) and the other deals with modeling and simulations issues(Section \ref{ss:path_modeling}). The two sets list several of the major issues that came up in our deliberations but they are by no means complete. In addition, we present, in Appendix~\ref{ss:path_approach}, a possible approach for addressing some of these problems and discuss it in the context of upcoming and future observational capabilities.

\subsection{Suggested Path Forward for Observations and Data Analysis} \label{ss:path_observations}

\noindent\textbf{Perform systematic observations of FC formation.}
In our deliberations, we realized that there exist only a handful of observations of FC formation, based mostly on the orientation of chromospheric fibrils \citep[e.g.,][]{gaizauskas1997,wang2007}. They concern relatively small FCs at the periphery of ARs, that form within 1--2 days. We could not find any study of the full timeline of formation of large quiescent FCs. The reasons may be that these FC are quite extended, form over long time-periods (weeks), and the existing high-resolution H$\alpha$ capabilities (needed to see fibril orientation) cannot cover the required spatial extensions and time periods. In principle, fibrils could be observed also in the EUV, but again one would need high resolution (IRIS), large FOV, and sufficiently long observation periods. Vector data could help to observe the development of shear, but the problem there is that most large FCs form in weak-field regions, where the field strengths are below the HMI sensitivity. 
Synoptic magnetic field and chromospheric observations from various viewpoints, i.e, terrestrial from SOT, HMI and DKIST and off the Sun-Earth line and above the ecliptic by Solar Orbiter's PHI, SPICE and EUI may allow the tracking the evolutionary paths of FC over longer time periods.

\noindent\textbf{Determine the role of cancellation in creating the pre-eruptive configuration.} Until now magnetic flux cancellation has been almost exclusively observed in regions where flux emergence has largely ceased. Flux cancellation could also occur in areas exhibiting  ongoing flux emergence, something that needs investigation, with the first steps towards this direction taken by \citet[][]{chintzoglou2019}.
In addition, the estimation of the magnetic flux associated with flux cancellation (Section \ref{sss:cance_obs}) and with coronal reconnection
estimated during confined flaring events 
\citep[e.g.,][]{veronig2015,tschernitz2018}
should clarify the respective role of flux cancellation, which is a continual and small-scale process, and coronal reconnection, which is a transient 
process with consequences for the large-scale structure
in establishing a pre-eruptive magnetic configuration.

\noindent\textbf{Perform statistical surveys for bodily emerging MFRs.} The observational cases in support of FC formation via  bodily/rigid MFR emergence are scarce. Moreover, the proposed interpretations are merely suggestive and inconclusive. To make progress, we need a comprehensive analysis of several cases. It is important to strictly consider FCs forming in isolation, as in \citet[][]{lites2010}, to strengthen the interpretations. Since flux emergence and FC formation may take more than a day, long-duration observing sequences will be necessary. Large coverage of the solar surface will make this exercise much easier, since the emergence may occur anywhere on the Sun and knowledge of the time history of the region is needed to decide whether the emergence occurs in a `filament-free' location. 

\noindent\textbf{Determine  the role of small-scale convective flows on the injection of helicity into the corona.} It is widely believed that convective flows at the photosphere are responsible for injecting the free energy that powers coronal heating, but their role in helicity injection has yet to be determined accurately. Vortical flows in inter-granular lanes are commonly observed, and the random nature of the convective flows may well impart a net twist to the coronal field,  eventually leading to the formation of FCs via helicity condensation, but this needs to be determined quantitatively. Observational studies are required that measure accurately the amount of helicity injected into the corona by the granular and supergranular motions. 

\noindent\textbf{Apply the SMA/MFR diagnostics matrix (Table\,\ref{t:table2}) to a statistically meaningful number of cases.} The maximum possible number of the observables of Table\,\ref{t:table2} should be considered. Such studies should provide a more coherent picture of the SMA/MFR signatures. Attention should be paid to 'hybrid' configurations, as they may be the norm. In particular, it should be examined how these structures evolve in time, and what is their typical state at eruption onset. In all cases, the properties (twist, helicity, magnetic flux, etc.) of each component  should be quantified.

\noindent\textbf{Benchmark the pre-eruptive magnetic configurations resulting from different NLFFF techniques.}
The properties of  modeled MFRs seem to depend on the particular NLFFF magnetic field extrapolations method, e.g  based on the widely-used \citet[][]{wiegelmann2014} code, or flux-rope insertion techniques \citep[][]{vanballegooijen04}, with the former typically leading to thinner MFRs \citep{LiuR&al2016}. These effects should be investigated in more detail, and should 
be compared also with results from magneto-frictional techniques \citep[e.g.,][]{mackay2011,fisher2015}, non-force-free field extrapolations \citep{HuQ&Dasgupta2008,zhu2016}, and MHD simulations \citep[e.g.,][]{torok18b}.

\noindent\textbf{Perform systematic studies of magnetic field in both quiescent and AR filaments.} 
The work on the IP and/or NP magnetic configurations stems from observational studies in the 1980s. A modern era study is needed to provide more complete and reliable observations. 

\noindent\textbf{Determine whether all sigmoids are signatures of MFRs.} Study more cases of sigmoid morphology and evolution in combination with coronal NLFFF and magnetofrictional modeling  \citep[as in][]{savcheva2009, savcheva2012,ZhaoJ&al2016, yardley2018b}, to clarify whether diffuse sigmoids already harbor an MFR and at what evolutionary stage.

\noindent\textbf{Determine the decay index of the coronal field at eruption onset.} Estimates of the decay index at the point of eruption onset for many events can help deciding between MHD instability models, which require an MFR to exist, and reconnection models, which operate in an SMA. 
A significant correlation between the estimated decay index at the observed onset height of eruptions
with the theoretically expected critical decay index for the  onset of the torus
instability found for a large sample of events supports the ideal MHD model,
i.e., the pre-existence of an MFR.
The reliability of present estimates \citep[e.g.,][]{ChengX&al2020} will be considerably improved when the precise tracking of eruptions near the limb with SDO/AIA data can be combined with simultaneous magnetograms of the source region from the Solar Orbiter in quadrature or from an L4 or L5 mission.

\subsection{Suggested Path Forward for Modeling and Simulations} \label{ss:path_modeling}
%
\noindent\textbf{Compare systematically the pre-eruptive configurations resulting from simulations of different FC formation mechanisms.} It is currently not possible to use the same boundary condition (same magnetogram) to inter-compare flux emergence, flux cancellation and helicity condensation simulations because they inherently change the photospheric boundary conditions as the models evolve. A systematic approach for this purpose is needed, but has not yet been developed. However, even exploiting existing simulations to compare the evolution and properties of pre-eruptive configurations produced by different mechanisms will lead to useful insights, as long as it is done in a systematic fashion.

\noindent\textbf{Perform comparative studies of full and partial flux-tube emergence.} 
A comparative study between the full (bodily) and partial emergence seen in flux-emergence simulations is sorely needed, aiming specifically to identify photospheric observational signatures that would allow one to distinguish between the two cases. If such signatures exist,  bodily-emerging MFRs could be identified, allowing their occurrence rate to be quantified. Such information would also provide valuable insight on the properties of flux tubes in the solar interior. Another important issue is to determine 
under which conditions a bodily-emerged MFR can erupt. Furthermore, the differences between the properties of bodily-emerged and post-emergence MFRs, formed by reconnection, have not yet been studied.

\noindent\textbf{Origins of photospheric shear.} 
Photospheric motions, such as shearing and rotation, which are driven by the Lorentz force when e.g. emerging twisted field expands into the corona, have long been studied for their role in the formation of MFRs and SMAs. In addition, the diverging motion of the polarities, which follows the emergence of the flux tube can act as a source of shearing at the photosphere.
We must systematically quantify the flows and
investigate their physical origins.

\noindent\textbf{Improve realism of flux-emergence simulations.}  
More realistic models of the flux-emergence process are crucial for the study of pre-eruptive structures. They are necessary in order to understand the role of flux cancellation in the formation of pre-eruptive structures and its coupling with magnetoconvection. The proper treatment of the physics of the lower solar atmosphere, where current models predict that MFRs initially form, are crucial to assess both the topology of the pre-eruptive strucures and also their mass and energy contents. In addition, the realistic treatment of the energy equation and the correct treatment of ion population are needed to accurately forward model the numerical simulations and compare with observations. 

\noindent\textbf{Determine the FC magnetic structure that results from helicity condensation.}
The amount of twist that can accumulate in an FC determines to a large degree its structural properties. Simulations of helicity condensation, to date, invariably resulted in an SMA configuration, but these studies were not performed at the observed scales for coronal helicity injection. It is unclear whether this result will hold in more rigorous calculations, or some type of MFR will be formed instead, and whether the helicity condensation process can account for the observed properties of FCs (e.g., their size and amount of shear/twist).

\noindent\textbf{Determine degree of MFR coherence to allow onset of eruption.} Perform data-constrained and data-driven numerical modeling of eruptions using MFRs with different degrees of coherence as initial condition. Such MFRs have been found in extrapolated NLFFFs and can also be constructed using the flux-rope insertion method or analytic MFR models \citep[e.g.,][]{titov14,titov18}. Initial conditions obtained with the novel non-force free extrapolation methods should also be considered.

\begin{acknowledgements}

We acknowledge the International Space Science Institute (ISSI) in Bern Switzerland for their generous support for travel and accommodations.
The authors would like to thank the two
anonymous referees for their very useful comments and
suggestions on the manuscript.
A.V. was supported by NASA grant NNX16AH70G and the LWS program through NNX15AT42G under ROSES NNH13ZDA001N. B.K. acknowledges useful discussions with Antonia Savcheva and Nishu Karna and support by the DFG and by NASA under Grants 80NSSC17K0016, 80NSSC18K1705, 80NSSC19K0082, and 80NSSC19K0860. J.Z. is supported by NASA grant NNH17ZDA001N-HSWO2R. J.E.L. acknowledges support by the NASA Living With a Star \& Solar and Heliospheric Physics programs, and the Office of Naval Research 6.1 Program and by the NRL-\textit{Hinode} analysis Program; the simulations were performed under a grant of computer time from the DoD HPC program. S.L.Y. would like to acknowledge STFC for support via the Consolidated Grant SMC1/YST025 and for a PhD Studentship. T.T. was supported by NASA's HGI and HSR programs (awards 80NSSC19K0263 and 80NSSC19K0858) and by NSF's PREEVENTS and Solar Terrestrial programs (awards ICER-1854790 and AGS-1923365). X.C. is funded by NSFC grants 11722325, 11733003, 11790303, Jiangsu NSF grants BK20170011, and “Dengfeng B” program of Nanjing University. V.A. and P.S. acknowledge support by the ERC synergy grant “The Whole Sun”. S.P. and A.V. would like to thank L. Vlahos for suggesting to establish an ISSI team. 

\end{acknowledgements}

\appendix
\appendixpage
\section{A Suggested Approach Towards Deciphering the Pre-Eruption Configuration} \label{ss:path_approach}
%
We suggest that \textit{vector magnetic field measurements at multiple heights spanning from the photosphere to the lower corona\/} and \textit{multi-viewpoint hot plasma observations in the corona\/} could advance our understanding of pre-eruptive configurations and may eventually lead to predicting their eruptions. To demonstrate the potential of such observations, we create and analyze synthetic data from an MHD flux-emergence simulation, as follows.

\begin{figure*}[h]
\centering
\includegraphics[scale=0.25]{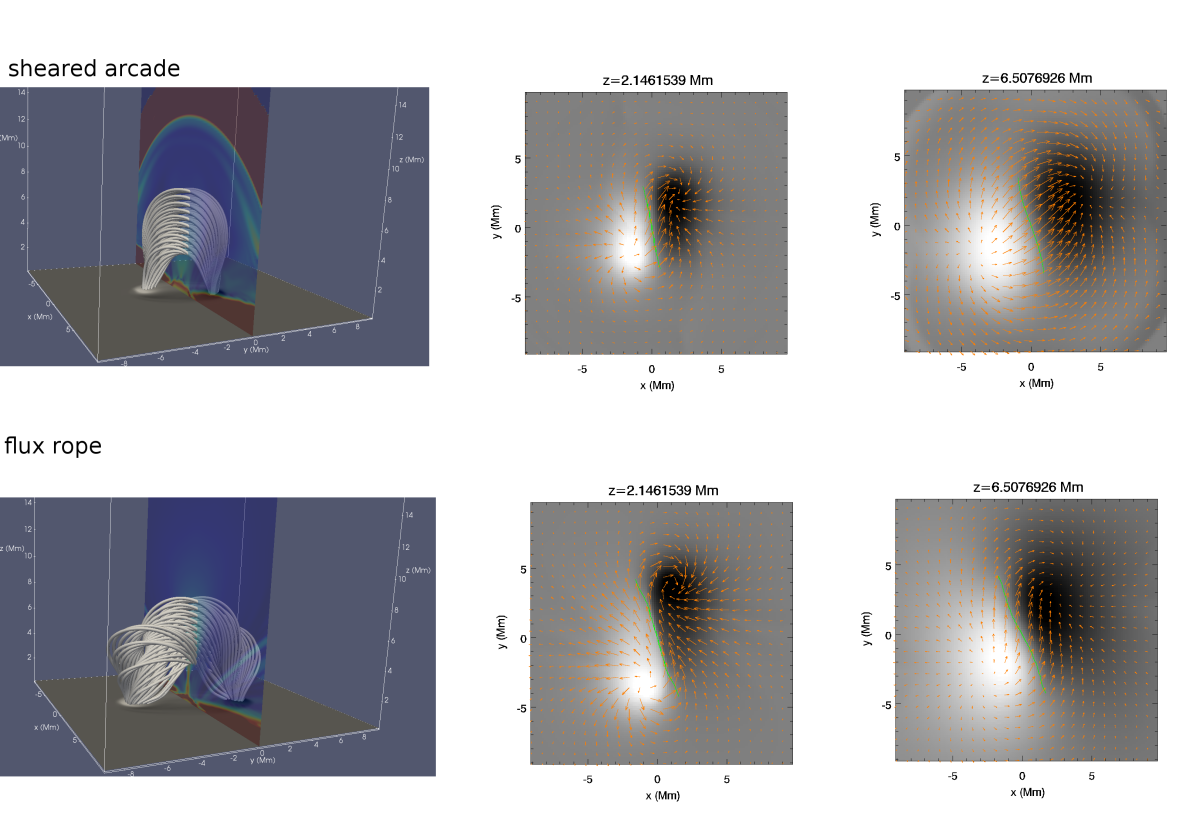}
\caption{ Two stages of a flux-emergence simulation  \citep{Syntelis_etal2017}, showing a simple SMA (top) that  transitions over time into an MFR (bottom). Left: representative magnetic field lines traced from identical seed points; $B_z$ is shown in greyscale at the photospheric level ($z=0$); the transparent vertical plane shows $|\mathbf{J}|/|\mathbf{B}|$. The middle and right columns show the PIL (green line) and $B_h$ (orange vectors) superposed on $B_z$ at z=2.14 Mm and z=6.5 Mm, respectively.}
\label{fig13}
\end{figure*}

First, we extract data-cubes of the vector magnetic field in the lower solar atmosphere from the MHD flux-emergence simulation of \cite{Syntelis_etal2017} (SAT17, hereafter). Note that the SAT17 simulation had a different objective. It was not meant to simulate a large-scale CME but it has all the necessary characteristics to demonstrate our approach. We select representative pre-eruptive configurations from this simulation at two consecutive times (Fig.\,\ref{fig13}). The configuration at the earlier time (top panels) consists predominantly of sheared, high-arching field lines of similar orientation, resembling a `2.5D-like' SMA that has a much simpler structure than the flat 3D SMAs described in Section\,\ref{ss:SMA}. At the later time (bottom panels), the configuration contains also twisted field lines, i.e., it has transitioned into an MFR.
The central and right panels show the horizontal magnetic field, $B_h$, at two different heights (corresponding roughly to the lower chromosphere and middle transition region).
The orientation of $B_h$ with respect to the PIL
is significantly different for the two magnetic configurations. For the SMA, $B_h$ points roughly along the same direction in both atmospheric layers. For the MFR, B$_h$ is much more sheared and flips orientation across the PIL with increasing height, which suggests a transition across the center of an MFR. This suggests that mapping the orientation of B$_h$ as a function of height in the low solar atmosphere may be a powerful approach for assessing the evolutionary stage (SMA-to-MFR transition) of pre-eruptive configurations. The height of this transition will depend on the physical parameters of a given region on the Sun. For this reason, multiple height coverage extending to coronal heights will be needed to properly capture the evolution of the system across the range of ARs and phases of solar activity. As a side benefit, the use of constraints from multiple atmospheric layers will greatly improve the quality of NLFFF (or non-force free) extrapolations and should lead to a more accurate specification of the magnetic field in the corona. 

\begin{figure*}[h]
\centering
\includegraphics[scale=0.2]{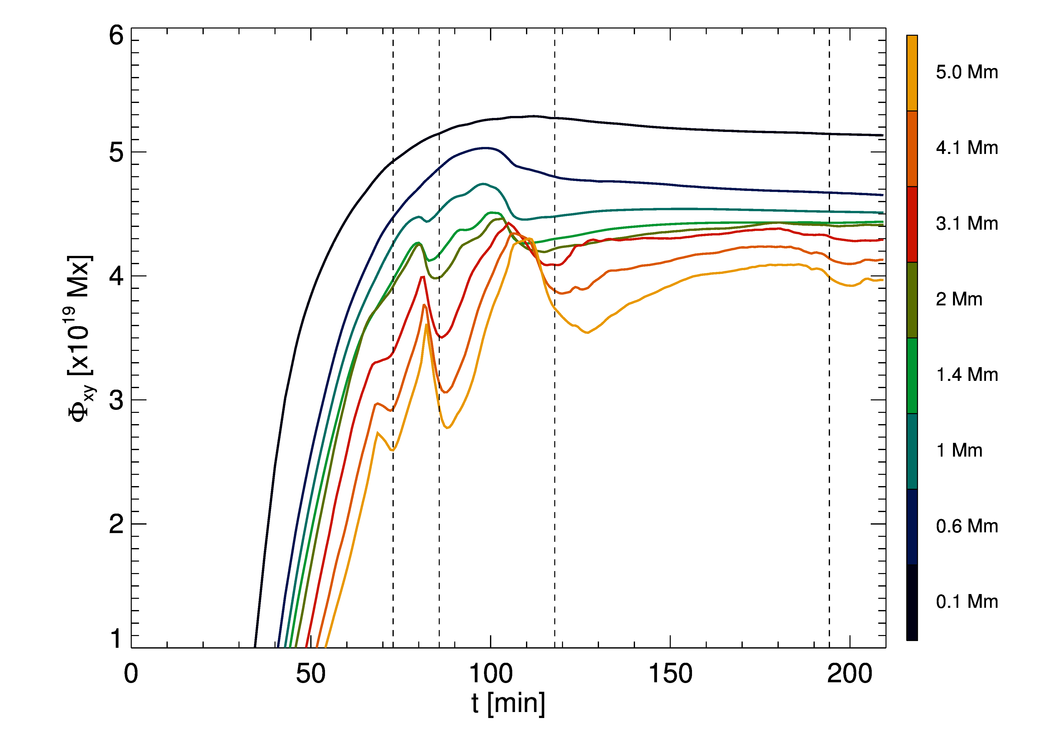}
\caption{Temporal evolution of magnetic flux at various heights for the \cite{Syntelis_etal2017} simulation. The snapshots employed in Figures \ref{fig13} and \ref{fig15} are taken before t=50 min. The vertical dashed lines show the kinetic energy maxima inside the numerical domain indicating the four eruptions in the simulation. }
\label{fig14}
\end{figure*}

Magnetic field observations above the photosphere should also capture  field changes induced by eruptions. Such changes have been notoriously difficult to detect in photospheric magnetograms \citep{sudol2005}, leading to uncertainties of the amount of magnetic flux removed by CMEs. We demonstrate this by exploiting the recurrent eruption characteristic of the SAT17 simulation in Figure \ref{fig14}. We see that while the photospheric magnetic flux exhibits a smooth behavior characterized by a sharp increase followed by a plateau (black line), the magnetic flux higher up undergoes several dips, each associated with an eruption (colored lines). It is, therefore, conceivable, that we could measure the flux (and by extension, the magnetic energy) that participates in an eruption by following the magnetic flux evolution above the photosphere. Constraining the energetics of eruptions with such direct measurements would greatly advance our understanding of explosive energy release and help establish the geo-effective potential of CMEs at a very early stage of the eruption process \citep{vourlidas2019}.

\begin{figure*}[ht]
\centering
\includegraphics[scale=0.2]{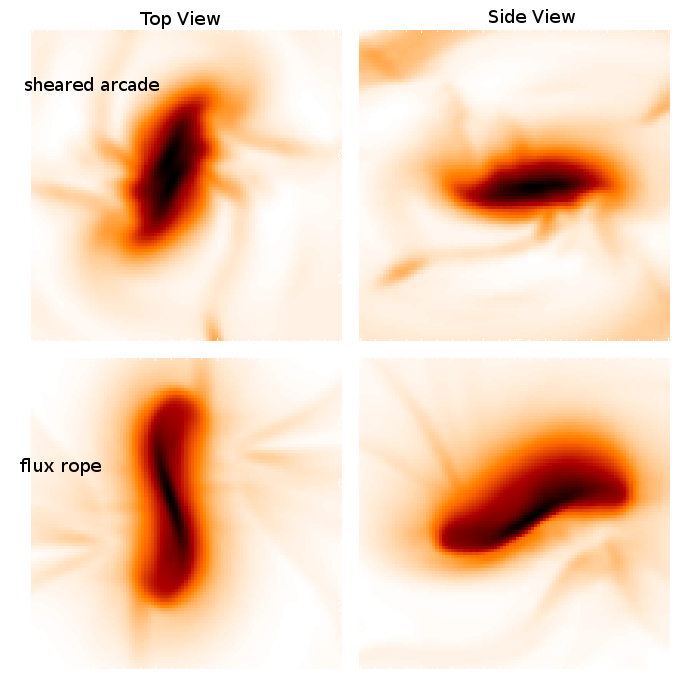}
\caption{LOS-integrations of ${J}^2$ for the 
two simulation snapshots of Figure \ref{fig13}. The left panels correspond to top views of the SMA and MFR in Figure \ref{fig13}. The right panels correspond to side views of the SMA and MFR. A reverse color-table is used.}
\label{fig15}
\end{figure*}

Finally, we calculate synthetic images resulting from LOS-integration of 
the squared electric current density, $J^2$, (Figure \ref{fig15}). These images can be considered as proxies for high-temperature plasma emissions in the EUV or SXR. We use the same simulation snapshots as in Figure \ref{fig13}. The top view of the SMA (Fig.~\ref{fig15}, top left) is reminiscent of a sigmoidal structure. This is, however, easily dismissed by inspecting the side view (top right). On the other hand, both views of the MFR snapshot display a clear sigmoidal structure. We emphasize that this is just a proof-of-concept attempt. It should be expanded with more rigorous calculations by, for instance, calculating synthetic Stokes profiles from the MHD simulations and inverting them as done with real data.

These diagnostics can be exploited once vector magnetic field observations at several layers above the photosphere become available. Photospheric and chromospheric vector magnetic field observations (albeit over a limited FOV) will be available from DKIST, beginning in late 2020. 
In addition, the Cryogenic Near Infra-Red Spectro-Polarimeter (Cryo-NIRSP) of DKIST will obtain off-limb observations in HeI 10830 \AA\ \, and FeXII 10747 \AA\ for estimates of magnetic field magnitudes and orientation in prominences and coronal loops. Even though these are not vector magnetograms, the polarization information along with forward modeling can be used to decipher the nature of pre-eruptive structures within coronal cavities as shown by \citet{Rachmeler&al2013}.
In late 2021, the remote sensing instruments on Solar Orbiter \citep{solo1,solo2} will start science operations, including photospheric vector magnetic field measurements by the Polarimetric and Helioseismic Imager (PHI) from non-Earth viewpoints, thus providing additional constraints on the photospheric boundary conditions for magnetic-field extrapolations and possibly allowing concurrent coronal and photospheric magnetic field estimates when combined with Cryo-NIRSP.  

Multi-viewpoint observations of hot plasmas are less certain in the near future, although some research can be undertaken either with past (2007-2014) EUVI 284 \AA\ observations from STEREO-A and -B  or now with comparisons between STEREO/EUVI 284 \AA\ (currently at L5 and approaching Earth) and SUVI 284\AA\ observations. In addition, the off-ecliptic viewpoints from Solar Orbiter will supply complementary high-resolution views of prominences and hot plasmas from the EUI and SPICE instruments, which could be compared with Earth-based  observations (e.g., SDO/AIA and SUVI). 

Further in the future, we are looking forward to exciting mission and instrument concepts. The proposed COSMO observatory \citep{Tomczyk&al2016} will derive some coronal magnetic field quantities (e.g strength, azimuth) and plasma properties further from the limb than DKIST. The SOLAR-C mission, recently approved by JAXA, will carry a next-generation upper atmospheric imaging spectrograph that will greatly enhance our diagnostics of the thermal structure  and the dynamics of pre-eruptive configurations. Based on our earlier discussion, an EUV channel similar to the 131 \AA \, channel of SDO/AIA, at an off Sun-Earth viewpoint, along with a vector magnetograph would make great addition to a scientific payload for an L5 mission \citep[][]{vourlidas2015}. An observatory at that location could play a major role in understanding CME initiation because it would increase the observational coverage of the solar surface magnetic field and the identification and tracking of magnetic regions would be possible for longer times \citep[e.g.,][]{mackay2016}. All these observations combined with a well-crafted modeling program will allow us to finally fully characterize the nature and evolution of pre-eruptive configurations towards eruption.

%

             \bibliographystyle{spbasic} 

\bibliography{issi.bib}


\end{document}